\documentclass[12pt]{article} 
\pdfoutput=1
\usepackage{graphicx,floatflt,amssymb,rotate}
\usepackage{mathrsfs}
\usepackage{amssymb}
\usepackage{amsmath}
\usepackage{color}
\usepackage{graphicx}
\usepackage{subfigure}
\usepackage{multirow}
\usepackage{float}
\usepackage{pstricks}
\usepackage[toc,page]{appendix}
\usepackage[mathscr]{euscript}
\usepackage{cite}
\usepackage[colorlinks=true,
linkcolor=red,
urlcolor=blue,
citecolor=blue]{hyperref}
\usepackage{color}

\newcommand{\mm}{\mathcal{M}}
\newcommand{\diag}{\mbox{diag}\,}

\def\q2 {q^2}

\def\bt{\begin{table}}
\def\et{\end{table}}

\def\mET{E_T \hspace{-1.0em}/\;\:}

\begin{document}

\vskip 30pt 

\begin{center}
	{\Large \bf Distinctive Collider Signals for a Two Higgs Triplet
          Model } \\
		\vspace*{1cm} {\sf ~Dilip Kumar
                  Ghosh$^{a,}$\footnote{tpdkg@iacs.res.in}, ~Nivedita
                  Ghosh$^{a,}$\footnote{tpng@iacs.res.in
                    (Corresponding author)}, ~Biswarup
                  Mukhopadhyaya$^{b,}$\footnote{biswarup@hri.res.in}}\\
		\vspace{10pt} {\small } {\em $^a$School Of Physical Sciences, Indian Association for the
                  Cultivation of Science,\\ 2A $\&$ 2B, Raja
                  S.C. Mullick Road, Kolkata 700032, India \\ $^b$
                  Regional Centre for Accelerator-based Particle
                  Physics \\ Harish-Chandra Research Institute}
\end{center}

\begin{abstract}
 \noindent 
The extension of the Standard Model (SM) with two complex $SU(2)_{L}$
scalar triplets enables one to have the Type II seesaw mechanism
operative consistently with texture-zero neutrino mass matrices.  This
framework predicts additional doubly charged, singly charged and
neutral spinless states. We show that, for certain values of the model
parameters, there is sufficient mass splitting between the two doubly
charged states ( $H_1^{\pm\pm}, H_2^{\pm\pm}$) that allows the decay
$H_1^{\pm\pm} \to H_2^{\pm\pm} h $, and thus leads to a unique
signature of this scenario. We show that the final state $2(\ell^{\pm}
\ell^{\pm}) + 4b + \mET~$ arising from this mode can be observed at
the high energy, high luminosity (HE-HL) run of the 14 TeV Large
Hadron Collider (LHC), and also at a 100 TeV Future Circular
Collider (FCC-hh).
 
 \vskip 5pt \noindent 
\end{abstract}


\renewcommand{\thesection}{\Roman{section}}  
\setcounter{footnote}{0}  
\renewcommand{\thefootnote}{\arabic{footnote}}  


\section{Introduction}
A 125 GeV scalar, with a striking resemblance to the Higgs boson
proposed in the Standard Model(SM) has been observed at the Large
Hadron Collider(LHC) ~\cite{Aad:2012tfa,Chatrchyan:2012xdj}. In spite
of being a very successful phenomenological theory, the SM, however,
cannot generate neutrino masses as suggested in various observations~\cite{Walter:2008ys, GonzalezGarcia:2007ib, Mohapatra:2006gs}. A
popular set of mechanisms for generating such masses are the three
types of seesaw mechanism~\cite{Minkowski:1977sc,Yanagida:1979as,Mohapatra:1979ia,Schechter:1980gr,Magg:1980ut,
Cheng:1980qt,Lazarides:1980nt,Mohapatra:1980yp,Foot:1988aq,Franceschini:2008pz}.
Their experimental confirmation, on accelerators in particular are also of
considerable interest~\cite{delAguila:2008cj,Arhrib:2009mz,Akeroyd:2005gt, Akeroyd:2009hb, Akeroyd:2010je,Akeroyd:2011zza}.  Out of the three suggested types of
seesaw, Type II involves an extension of the scalar sector with an
additional complex $SU(2)_L$ triplet scalar with hypercharge $Y = 2$. This triplet
couples to leptons via interactions which violate lepton number by two
units~\cite{Schechter:1980gr,Magg:1980ut,Cheng:1980qt,Lazarides:1980nt,Mohapatra:1980yp,
  Lindner:2016bgg, Arhrib:2011uy} and thus generates Majorana masses
for neutrinos.

The most striking phenomenological consequence of such a triplet
scalar is the presence of a doubly charged scalar. The triplet vacuum
expectation value (vev), denoted here by $w$, is rather tightly
constrained by the $\rho$-parameter to values less than about 5
GeV~\cite{Aoki:2012jj}.  The doubly charged scalar can decay either to
produce same-sign dilepton peaks for $w < 10^{-4} ~\rm GeV$, or to
same-sign $W$-pairs for $w > 10^{-4} ~\rm
GeV$~\cite{Perez:2008ha,Melfo:2011nx,Aoki:2011pz}.  A rather strong
lower limit of 770 - 800 GeV exists on the doubly charged scalar mass
in the former case, from same-sign dilepton searches at the LHC
~\cite{ATLAS:2017iqw}.  There is no such bound yet on doubly charged
scalar mass for $w > 10^{-4}~\rm GeV$ ~\cite{Han:2007bk}. This is
because (a) a relatively large triplet vev implies small $\Delta L =
2$ Yukawa couplings from a consideration of neutrino masses, and (b)
overcoming standard model (SM) backgrounds for the final state driven
by same-sign W-pairs is a challenging task, and requires a large
integrated luminosity.  Several works can be found in the literature,
dwelling on strategies for unraveling the triplet scenario, both
before ~\cite{Chakrabarti:1998qy,Cheung:2002gd,Akeroyd:2005gt, Akeroyd:2009hb, Akeroyd:2010je,
  Akeroyd:2011zza, Chiang:2012dk} and after \cite{Kanemura:2013vxa,
  kang:2014jia, Kanemura:2014goa,Kanemura:2014ipa,Chen:2014qda,
  Han:2015hba, Han:2015sca, Mitra:2016wpr,Ghosh:2017pxl,Dev:2018kpa} the discovery
of the 125 GeV scalar.

From some special angles, however, a single triplet is inadequate for
consistent neutrino mass generation in the Type-II seesaw model. For
example, the somewhat different mass and mixing patterns in the
neutrino sector (as compared to those in the quark sector) calls for
studies in neutrino mass matrix models. One class of such models
consists of zero textures, having some vanishing entries in the mass
matrix, thus leading to relations between mass eigenvalues and mixing
angles, and ensuring better predictiveness in the neutrino sector. It
has been shown, that zero textures are inconsistent with Type-II
seesaw models with a single scalar triplet~\cite{Grimus:2004hf}. Such inconsistency is
removed when one has two such triplets, as has been demonstrated in~\cite{Chaudhuri:2013xoa}. 
This of course opens up the possibility of new collider signals which has been only
partially investigated.

In this work, we study the LHC signals that can decidedly tell us
about the existence of two complex triplet
scalars($\Delta_1,\Delta_2$). For example, in 
~\cite{Chaudhuri:2013xoa} searches via $\Delta_1^{\pm\pm} \rightarrow
\Delta_2^{\pm} W^{\pm}$ decay mode have been discussed. It
should be noted that such a decay is disfavored in the
single-triplet scenario, since the $\rho$-parameter restricts the
mass splitting among fields having different electric charges.
Another decay channel that opens up in this scenario is
$\Delta_1^{\pm\pm} \rightarrow \Delta_2^{\pm\pm} h$, h being the SM-like 
Higgs.  As has been shown in~\cite{Chaudhuri:2016rwo}, this mode
prevails especially in the presence of at least one CP-violating
phase.
 
We neglect CP-violating phases in the present study. Regions of the
parameter space answering to both  $\Delta_1^{\pm\pm} \rightarrow
\Delta_2^{\pm} W^{\pm}$ and $\Delta_1^{\pm\pm} \rightarrow
\Delta_2^{\pm\pm} h$  have been identified, and the
corresponding signals have been predicted. Both these channels can
lead to the final state $ 2\ell^\pm \ell^\pm+ 4 b + \mET $, 
where, $\ell \equiv e, \mu $. We carry out a detailed analysis
to estimate signal significance of this scenario in the regions of the
parameter space, consistent with all current limits both at the high
energy high-luminosity (HE-HL) run of the LHC with $\sqrt{s} = 14$~TeV
and at the proposed $\sqrt{s} = 100 $ TeV Future Circular Collider
(FCC-hh) at CERN \cite{Golling:2016gvc} or the Super Proton-Proton
Collider (SPPC) in China \cite{Tang:2015qga}. Further details on the
physics potential of the 100 TeV collider can be found, for example, in
\cite{Arkani-Hamed:2015vfh}. We also comment on how to
differentiate the two-triplet scenario from a single-triplet one using
the signal analysis in this work.

The paper is organized as follows. In Section II we discuss a little bit
about the well-motivated $Y=2$ single triplet scenario. Relevant
phenomenological features of the two-triplet scenario are presented in Section III. 
Results of the collider analysis are reported in Section IV. We summarize and conclude 
in Section V.


\section{The $Y=2$ single triplet scenario}
In this section we briefly describe the single triplet scenario.
Along with the SM fields, there is an extra $SU(2)_L $ complex triplet
scalar field $\Delta$ with hypercharge $Y=2$.
\begin{eqnarray}
\Delta &=& \frac{\sigma^i}{\sqrt 2}\Delta_i = \left(\begin{array}{cc}
\delta^+/\sqrt 2 & \delta^{++}\\
\delta^0 & -\delta^+/\sqrt 2
\end{array}
\right),
\end{eqnarray}

where $\Delta_1=(\delta^{++}+\delta^0)/\sqrt
2,~\Delta_2=i(\delta^{++}-\delta^0)/\sqrt 2,~\Delta_3=\delta^+$.

The vevs of the doublet and the triplet
are given by
\begin{equation}\label{vev}
\langle \phi \rangle_0 =
\frac{1}{\sqrt{2}} \left(\begin{array}{c} 0 \\ v \end{array}\right)
\quad \mbox{and} \quad
\langle \Delta \rangle_0 =
\left(\begin{array}{cc} 0 & 0 \\ w & 0 \end{array}\right),
\end{equation}
respectively and the electroweak vev is given by $v = \sqrt{v^2 +
  2w^2} = 246 ~\rm GeV$.

The most general scalar potential involving $\phi$ and $\Delta$ can be
written as
\begin{eqnarray}
V(\phi,\Delta)\ = \hphantom{+} \rm a\, \phi^\dagger\phi + \ \frac{\rm
  b}{2} \rm tr(\Delta^\dagger\Delta)+ \ c\, (\phi^\dagger\phi)^2 +
\ \frac{\rm d}{4}\left( \rm tr(\Delta^\dagger\Delta) \right)^2
\nonumber \\ + \ \frac{\rm r}{2} \phi^\dagger\phi \rm
tr(\Delta^\dagger\Delta) + \ \frac{\rm f}{4} \rm
tr(\Delta^\dagger\Delta^\dagger)\rm tr(\Delta\Delta) \nonumber \\ +
\rm h \phi^\dagger \Delta^\dagger \Delta \phi + \left( t\,
\phi^\dagger \Delta \tilde{\phi} + \mbox{H.c.} \right),
\end{eqnarray}
where $\tilde{\phi} \equiv i\tau_2 \phi^\ast$.  In general, both $v$
and $w$ can be complex. However, since we want to avoid all
CP-violating effects, we choose both the vevs to be real and positive,
which as a result implies that $t$ has to be real.

It should be remembered that the choice $a<0$, $b>0$
ensures the primary source of spontaneous symmetry breaking
to be the vev of the scalar doublet. At the same time, the
$\rho$-parameter has to be very close to its tree-level SM value of 
unity, as required by the latest data, namely, $\rho =
1.0004^{+0.0003}_{-0.0004}$~\cite{Olive:2016xmw} for
$w \ll v$. Also the doublet-triplet mixing has to be small and 
the perturbativity of all quartic couplings at the electroweak scale 
has to be guaranteed.

All the aforementioned constraints drive us to choose the following orders of
magnitude for the parameters in the potential:
\begin{equation}\label{oomagn}
a,\: b \sim v^2; \quad
c,\: d,\: r,\: f,\: h \sim 1; \quad 
|t|\ll v.  
\end{equation}

The mass terms for  singly-charged scalars in this model are given by
\begin{equation}
\mathcal{L}^\pm_S = -
(H^-, \phi^-) \mathcal{M}_+^2 
\left( \begin{array}{c} H^+ \\ \phi^+ \end{array} \right)
\end{equation}
where
\begin{equation}
\mathcal{M}_+^2 = \left(\begin{array}{cc}
(q+h/2)v^2 & \sqrt{2} v (t - w h/2) \\
\sqrt{2} v (t - w h/2) & 2(q + h/2)w^2 
\end{array}\right) 
\quad \mbox{and} \quad 
q = \frac{|t|}{w}.
\end{equation}

Diagonalization of the matrix should yield
one zero eigenvalue, corresponding to the
Goldstone boson. The singly-charged mass-squared matrix 
is 
\begin{equation}
m^2_{\Delta^+} = \left(q + \frac{h}{2} \right) (v^2 + 2w^2),
\end{equation} 
whereas the doubly-charged scalar mass is
\begin{equation}
m^2_{\Delta^{++}} = (h + q) v^2 + 2fw^2.
\end{equation}
In the limit  $w \ll v$, one obtains 
\begin{equation}
m^2_{\Delta^{++}} - m^2_{\Delta^+} \simeq {\frac {h}{2}}v^2.
\end{equation}

Electroweak precision data imply $\Delta M \equiv \mid
m_{\Delta^{\pm \pm}} - m_{\Delta^\pm}\mid  \lesssim
50$ GeV~\cite{Chun:2012jw,Baak:2014ora} assuming a light SM Higgs
boson of mass $m_h = 125$ GeV and top quark mass $M_t = 173 $ GeV.
Hence, the decay mode $\Delta^{++} \rightarrow \Delta^+ W^+$ is
kinematically not allowed with a single triplet scalar.

\section{Extension with two triplets}
The single-triplet scenario can sometimes turn out to be inadequate.
For example, the somewhat novel kind of bi-large mixing in the neutrino
sector motivates people to link such a mixing pattern with the neutrino mass
matrix itself. The number of arbitrary parameters in such an investigation 
is reduced, and the mass eigenvalues and mixing angles are linked in a
predictive manner, if some elements of this matrix vanish. It is with this in view that     
various texture zero neutrino mass matrices have been proposed, for example,
through the imposition of certain Abelian symmetries. Two-zero
textures constitute a popular subset of such models, which have
been widely used in various contexts~\cite{Frampton:2002yf,Xing:2002ta,Xing:2002ap,Honda:2003pg,
Guo:2003cc,Grimus:2004hf,Honda:2004qh,Goswami:2008rt,Choubey:2008tb,Goswami:2009bd,Fritzsch:2011qv,Ghosh:2012pw,
Liao:2015hya,Kitabayashi:2015jdj,Lamprea:2016egz}.

In the specific context of Type II seesaw, however, inconsistencies
arise when texture zeros (especially two-zero textures) are
attempted~\cite{Grimus:2004az}. Such inconsistency can be avoided, as
already mentioned, when two scalar triplets are present. In such a
scenario, one extends the SM with two $Y=2$ triplet scalars
$\Delta_1$, $\Delta_2$:
\begin{equation}\label{triplet2}
\Delta_1 = 
\left(\begin{array}{cc}
\delta_1^+ & \sqrt{2}\delta_1^{++} \\ \sqrt{2}\delta_1^0 & -\delta_1^+ 
\end{array}\right)
\quad \mbox{and} \quad
\Delta_2 = 
\left(\begin{array}{cc}
\delta_2^+ & \sqrt{2}\delta_2^{++} \\ \sqrt{2}\delta_2^0 & -\delta_2^+ 
\end{array}\right).
\end{equation}
The vevs of the scalar triplets are given by
\begin{equation}\label{vev2'}
\langle \Delta_1 \rangle_0 =
\left(\begin{array}{cc} 0 & 0 \\ w_1 & 0 \end{array}\right) 
\quad \mbox{and} \quad
\langle \Delta_2 \rangle_0 =
\left(\begin{array}{cc} 0 & 0 \\ w_2 & 0 \end{array}\right).
\end{equation}

The scalar potential in this scenario involving the Higgs doublet and
the two triplets can be written as
\begin{eqnarray}
\lefteqn{V(\phi,\Delta_1,\Delta_2) =}
\nonumber \\
&&
\rm a\, \phi^\dagger\phi + 
\frac{1}{2}\,\rm b_{kl} \rm tr(\Delta_k^\dagger \Delta_l)+ 
\rm c (\phi^\dagger\phi)^2 + 
\frac{1}{4}\, \rm d_{kl} \left( \rm tr(\Delta_k^\dagger\Delta_l) \right)^2
\nonumber \\
&&
+ \frac{1}{2}\,\rm r_{kl} \,
\phi^\dagger \phi  \rm tr (\Delta_k^\dagger\Delta_l) +
\frac{1}{4}\,\rm f_{kl}  
\rm tr(\Delta_k^\dagger\Delta_l^\dagger) \rm tr(\Delta_k\Delta_l)
\nonumber \\
&&
+ \rm h_{kl}\, \phi^\dagger \Delta_k^\dagger \Delta_l \phi +
\rm g \rm tr(\Delta_1^\dagger\Delta_2) \rm tr(\Delta_2^\dagger\Delta_1) + 
\rm g' \rm tr(\Delta_1^\dagger\Delta_1) \rm tr(\Delta_2^\dagger\Delta_2) 
\nonumber \\
&&
+ \left( \rm t_k\, \phi^\dagger \Delta_k \tilde{\phi} +
\mbox{H.c.} \right),
\label{pot}
\end{eqnarray}
where $k,l=1,2$.

As mentioned in the previous section,
$v, w_1, w_2$ as well as $t_1, t_2$ are taken to be real and positive.  
One can also use for illustration, without any loss in the generality of the results,
\begin{equation}\label{oomagn2}
a,\: b_{kl} \sim v^2; \quad
c,\: d_{kl},\: r_{kl},\: h_{kl},\: f_{kl},\: g,\: g' \sim \mathcal{O}(1); \quad
|t_k| \ll v. 
\end{equation}
With $w_1, w_2 \ll v$.

We redefine the following $2 \times 2$ matrices and vectors: 
\begin{equation}\label{param_model} 
B = (b_{kl}), \quad E = (r_{kl}+ h_{kl}), \quad H = (h_{kl}), \quad
t = \left( \begin{array}{c} t_1 \\ t_2 \end{array} \right), \quad
w = \left( \begin{array}{c} w_1 \\ w_2 \end{array} \right).
\end{equation}

The minimization of the potential~(\ref{pot}), neglecting all terms quartic in the
triplet vevs, yields
\begin{eqnarray}
\left( B + \frac{v^2}{2} \left( E-H \right) \right) w + 
v^2\, t &=& 0, 
\label{vev1} \\
a + cv^2 + \frac{1}{2} w^T (E-H) w + 2\, t \cdot w &=& 0,
\label{vev2}
\end{eqnarray}
where we have used $t \cdot w = \sum_k t_k w_k$.
Solving Eq.~(\ref{vev1}) and Eq.~(\ref{vev2}) simultaneously,  $w_k (k=1,2)$
are obtained as
\begin{equation}
w = -v^2 \left( B + \frac{1}{2} v^2 (E-H) \right)^{-1} t.
\end{equation}

After diagoalization of different kinds of scalar mass matrices following 
electroweak symmetry breaking (EWSB), we obtain the
charged scalars ($H^{\pm \pm}_1,H^{\pm \pm}_2$), singly charged
Higgs($H^{\pm}_1,H^{\pm}_2$),  and the neutral CP-even($h,H_1,H_2$) 
and CP-odd($A_1,A_2$) scalars. Among them  $h$ is the SM-like Higgs.

The mass matrix of the doubly-charged scalars is given by
\begin{equation}
\mm^2_{\pm \pm} = B + \frac{v^2}{2} \left( E+H \right).
\end{equation}
which can be diagonalized by 
\begin{equation}
U^\dagger \mm^2_{\pm \pm} U = \diag (M^2_1, M^2_2)
\end{equation}
yielding the mass eigenstates:
\begin{equation}
\left( \begin{array}{c} \delta^{\pm \pm}_1 \\ \delta^{\pm \pm}_2
\end{array} \right) = U
\left( \begin{array}{c} H^{\pm \pm}_1 \\ H^{\pm \pm}_2
\end{array} \right)
\end{equation}

The singly charged scalar mass-squared matrix comes from 
\begin{equation}
-\mathcal{L}^{\pm}_S = 
\left( \delta^-_1, \delta^-_2, \phi^- \right) \mm^2_{\pm}
\left( \begin{array}{c}
\delta^+_1 \\ \delta^+_2 \\ \phi^+ \end{array} \right)
+ \mbox{H.c.},
\end{equation}

where
\begin{equation}\label{m+}
\renewcommand{\arraystretch}{1.2}
\mm^2_{\pm} = \left(
\begin{array}{cc}
B + \frac{v^2}{2} \, E & 
\sqrt{2} v \left( t - H w/2 \right) \\
\sqrt{2} v \left( t - H w/2 \right)^\dagger &
a + cv^2 + \frac{1}{2} w^T (E+H) w
\end{array} \right).
\end{equation}
Using equations~(\ref{vev1}) and~(\ref{vev2}) we get, 
\begin{equation}\label{wouldbe}
\mm^2_{\pm} \left( \begin{array}{c}
v_T \\ v/\sqrt{2} \end{array} \right) = 0.
\end{equation}
This serves as a consistency check that the singly charged mass
matrix has to have an eigenvector with zero eigenvalue that
corresponds to the would-be-Goldstone boson.

It is evident from Eq.~(\ref{oomagn2}) that b is of the order of $v^2$. Therefore, 
in a rough approximation one can safely ignore the $t_k$ and the triplet vevs in
the mass matrix $\mm^2_{\pm}$. In that limit, also $a + cv^2 = 0$ and 
the charged would-be-Goldstone boson is equivalent to $\phi^{\pm}$. There is no 
mixing with the $\delta^{\pm}_k$. 

The singly charged mass matrix can be diagonalized by 
\begin{equation}
V^\dagger \mm^2_{\pm} V = \diag (\mu_1^2, \mu_2^2, 0),
\end{equation}
with
\begin{equation}
\left( \begin{array}{c} \delta^{\pm}_1 \\ \delta^{\pm}_2 \\ \phi^+
\end{array} \right) = V
\left( \begin{array}{c} H^{\pm}_1 \\ H^{\pm}_2 \\ G^{\pm}
\end{array} \right).
\end{equation}
Where $G^{\pm}$ is the charged would-be-Goldstone boson.

Interactions with the W-boson are given by 
\begin{eqnarray}
\mathcal{L}_\mathrm{gauge} &=& ig \sum_{k=1}^2 \left[
\delta^-_k (\partial^\mu \delta^{++}_k) -
(\partial^\mu \delta^-_k)\, \delta^{++}_k  
\right] W_\mu^- 
\nonumber \\ 
&& 
-\frac{g^2}{\sqrt{2}} \sum_{k=1}^2 w_k W^-_\mu {W^-}^\mu \delta_k^{++}
+ \mbox{H.c.} 
\end{eqnarray}
Here $g$ is the $SU(2)_{L}$ gauge coupling constant.
Changing the gauge basis into mass basis 
allows us to compute the decay rates of 
$H_1^{++} \to H_2^+ W^+$ and $H_k^{++} \to W^+ W^+$ ($k=1,2$).

The ($\Delta L = 2$) Yukawa interaction Lagrangian involving the
triplets and the leptons is
\begin{equation}\label{Ll}
\mathcal{L}_Y = \frac{1}{2} \,
\sum_{k=1}^2 h^{(k)}_{ij} L^T_i C^{-1} i \tau_2 \Delta_k L_j + 
\mbox{H.c.},
\end{equation}
where $L_i$
denote the left-handed lepton doublets, $C$ is the Dirac charge conjugation matrix, the $h^{(k)}_{ij}$ 
are the symmetric neutrino Yukawa coupling matrices of the triplets
$\Delta_k$, and the $i,j = 1,2,3$ are the
summation indices over the three neutrino flavours.\footnote{We
  assume the charged-lepton mass matrix to be already
  diagonal.}

When the triplets acquire vev from Eq.~(\ref{Ll}) one can generate the neutrino mass matrix as :
\begin{equation}\label{ymat}
(M_\nu)_{ij}
= h^{(1)}_{ij} w_1 + h^{(2)}_{ij} w_2.
\end{equation}
This connects the Yukawa coupling constants $h^{(1)}_{ij}$, 
$h^{(2)}_{ij}$ and the triplet vevs $w_1$, $w_2$. 

In this work we use, once more as illustration, the normal hierarchy
of the neutrino mass spectrum and set the lowest neutrino mass
eigenvalue to zero.  The elements of The neutrino mass matrix $M_\nu$
can be obtained by using the observed central values of the various
lepton mixing angles and by diagonalising as
\begin{equation}\label{Mnu}
M_\nu = U {\hat{M}}_\nu U^\dagger,
\end{equation} 
where $U$ is the PMNS matrix given by~\cite{Beringer:1900zz} 
\begin{equation}
U = 
\left(\begin{array}{ccc}
c_{12} c_{13} & s_{12} c_{13} & s_{13} e^{-i\delta}  \\ - s_{12} c_{23} -
c_{12} s_{23} s_{13} e^{i\delta}  & c_{12} c_{23} - s_{12} s_{23} s_{13}
e^{i\delta} & s_{23} c_{13}  \\ s_{12} s_{23} - c_{12} c_{23} s_{13}
e^{i\delta} & - c_{12} s_{23} - s_{12} c_{23} s_{13} e^{i\delta} & c_{23}
c_{13} 
\end{array}\right)
\end{equation}
and ${\hat{M}}_\nu$ is the diagonal matrix of the neutrino masses.  We
have dropped possible Majorana phases for simplicity. Global analyses
of data can be used to resolve the various entries of
$U$~\cite{Fogli:2012ua}.  The left-hand side of Eq.~(\ref{ymat})
is reliably represented, at least in orders of magnitude, by the
central values of all angles, including that for $\theta_{13}$ as
obtained from the recent Daya Bay and RENO
experiments~\cite{An:2012eh,Ahn:2012nd}.  The actual mass matrix thus
constructed has some elements at least one order of magnitude smaller
than the others, thus suggesting texture zeros.


\section{Analysis}

Let us now look for smoking gun collider signals of doubly charged
scalars of this scenario with $w_k \sim {\cal O}(1) $ GeV. The
spectacular $l^{\pm}l^{\pm}$ decay channels are suppressed in this case.
The doubly charged scalars now mainly decay
into the following final states:
\begin{eqnarray}
H^{\pm\pm}_1 &\rightarrow& H^{\pm\pm}_2 h,\label{h} \\
H^{\pm\pm}_1 &\rightarrow& W^\pm W^\pm, \\
H^{\pm\pm}_1 &\rightarrow& H^\pm_2  W^\pm,\label{h2w}  \\
H^{\pm\pm}_2 &\rightarrow& W^\pm W^\pm.
\end{eqnarray}

The decay mode as mentioned in Eq.(~\ref{h}) is absent in the single-triplet model, since
there is only one doubly charged scalar particle.  Also, the
equivalent of Eq.(~\ref{h2w}), namely, $H^{\pm\pm}_1 \rightarrow H^\pm_1 W^\pm$ is kinematically disfavored since the mass splitting
between singly and doubly charged scalar is restricted by the $\rho$ parameter
constraint~\cite{Olive:2016xmw,Chun:2012jw,Baak:2014ora}. Hence, in order to
distinguish between the single triplet and the double triplet scalar
model, it is advantageous to investigate 
channels in Eq.(~\ref{h}) and Eq.(~\ref{h2w}), since the corresponding event topologies
cannot be faked by a single-triplet scenario. The production of  $W^\pm$  in association with the
SM-like Higgs boson leads to the following final state:
\begin{equation}
 2\ell^\pm \ell^\pm+ 4 b + \mET.\label{channel}
\end{equation}
where $\ell = e, \mu$, which arise from $W^\pm \to \ell^\pm \nu_\ell
({\bar \nu_\ell}) $ and $h \to b {\bar b}$ decay modes.

Since the doublet-triplet mixing is small in this model, there is no
noticeable difference in production rate in the gluon fusion channel
and also the tree-level decay of the SM-like Higgs.  However, the
presence of $H^{\pm \pm}_1,H^{\pm \pm}_2$ and $H^{\pm}_1,H^{\pm}_2$
modify the loop-induced $h\to \gamma \gamma$ decay
significantly. Detailed analyses of such modification can be found in
~\cite{Chun:2012jw,Aoki:2012jj,Arbabifar:2012bd,Kanemura:2012rs,
  Akeroyd:2012ms,Dev:2013ff,Das:2016bir}. Here we just ensure that
our benchmark points are consistent with the limits on the diphoton
signal strength of the Higgs at the $2\sigma$ level~\cite{Sirunyan:2018ouh}.

\subsection{Benchmark Points}
For collider analysis we choose three representative benchmark 
points such that the decays $H^{++}_1 \rightarrow H^{++}_2 h$ and
$H^{++}_1 \rightarrow H^{+}_2 W^{+}$ are kinematically allowed. Also, the mass splitting
$\Delta M $ between the doubly and singly charged higgses for 
each of the triplet scalar fields are consistent 
$( \Delta M \leq 50 ~\rm GeV)$~\cite{Chun:2012jw,Baak:2014ora} with the 
$\rho$ parameter constraint. All three benchmark points are also consistent
with the observed Higgs signal strength.  
In Table ~\ref{table:fixedparam} we show thirteen model parameters as defined
in the scalar potential in Eq.~\ref{pot}) that determine the benchmark points.
One can see from the table that the values of some of the parameters are kept 
fixed while others $(B,D,E,F,c)$ have been varied, since  
the scalar masses are strongly dependent on the latter\footnote{ We have kept the triplet vevs
fixed at $\sim 1$ GeV throughout our analysis to allow maximum mixing between the doublet and the triplet scalars as permitted
by $\rho$ parameter constraint~\cite{Aoki:2012jj}
for all the three benchmark points noted in Table~\ref{table:fixedparam}. In principle, one can reduce 
the vevs up to $10^{-4}$ GeV~\cite{Perez:2008ha,Melfo:2011nx,Aoki:2011pz} and that will not change the decay
modes of the non-standard scalars.}.
Table~\ref{table:benchmark} lists the corresponding values of neutral, 
singly charged and doubly charged scalar masses. The corresponding mixing angles in different sectors are
shown in Table~\ref{table:mixing angles}. The mixing angles between the doubly charged higgses, singly charged scalars,
and CP even neutral scalars (same as CP odd scalars) are denoted by 
$\alpha$,  $\beta$ and $\gamma$ respectively.

\begin{table}[ht]
\resizebox{18cm}{!}{
 \begin{tabular}{|c|c|c|c|c|c|c|c|c|c|c|c|c|c|c|}
  \hline
 & & a & B & D & E & H & F & c & g & $g^{'}$ & $t_{1}$ & $t_{2}$ & $w_{1}$ & $w_{2}$ \\
 & & & & & & & & & & & & & (in GeV) & (in GeV)\\
  \hline
 & BP1 & -15625 & $\left(\begin{array}{cc}
41012.0 & -31980.0\\
-31980.0 & 21182.4
\end{array}
\right)$ & $\left(\begin{array}{cc}
4.00 & 3.56\\
3.56 & 4.00 
\end{array}
\right)$ & $\left(\begin{array}{cc}
2.64 & 2.80\\
2.80 & 2.64
\end{array}
\right)$ & $\left(\begin{array}{cc}
1.00 & 1.00\\
1.00 & 1.00
\end{array}
\right)$ & $\left(\begin{array}{cc}
4.00 & 2.00\\
2.00 & 4.00
\end{array}
\right)$ & 0.26 & 0.90 & 0.90 &-1 & -2 & 1.09 & 1.32 \\
\hline
& BP2 & -15625 & $\left(\begin{array}{cc}
43012.0 & -31998.0\\
-31998.0 & 23182.4
\end{array}
\right)$ & $\left(\begin{array}{cc}
4.00 & 3.40\\
3.40 & 4.00 
\end{array}
\right)$ & $\left(\begin{array}{cc}
2.54 & 2.80\\
2.80 & 2.54
\end{array}
\right)$ & $\left(\begin{array}{cc}
1.00 & 1.00\\
1.00 & 1.00
\end{array}
\right)$ & $\left(\begin{array}{cc}
3.50 & 2.00\\
2.00 & 3.50
\end{array}
\right)$ & 0.24 & 0.90 & 0.90 &-1 & -2 & 1.09 & 1.32 \\
\hline
& BP3 & -15625 & $\left(\begin{array}{cc}
41012.0 & -31998.0\\
-31998.0 & 21182.4
\end{array}
\right)$ & $\left(\begin{array}{cc}
4.00 & 3.60\\
3.60 & 4.00 
\end{array}
\right)$ & $\left(\begin{array}{cc}
2.74 & 2.80\\
2.80 & 2.74
\end{array}
\right)$ & $\left(\begin{array}{cc}
1.00 & 1.00\\
1.00 & 1.00
\end{array}
\right)$ & $\left(\begin{array}{cc}
3.00 & 2.00\\
2.00 & 3.00
\end{array}
\right)$ & 0.25 & 0.90 & 0.90 &-1 & -2 & 1.09 & 1.32 \\
\hline
 \end{tabular}}
\caption{\it Parameters for all benchmark points.}
    \label{table:fixedparam}
\end{table}

\begin{table}
\centering
 \begin{tabular}{|c|c|c|c|c|c|c|c|c|}
  \hline
&  & $m_{H_1^{\pm\pm}}$ &  $m_{H_2^{\pm\pm}}$ &  $m_{H_1^{\pm}}$ & $m_{H_2^{\pm}}$ &  $m_{H_1}/m_{A_1}$ & $m_{H_2}/m_{A_2}$ & $\mu_{\gamma\gamma}$\\
&  & (in GeV) &  (in GeV) & (in GeV) & (in GeV) & (in GeV) & (in GeV) & \\
  \hline
 & BP1 & 416.37 & 216.13 & 402.13 & 215.87 & 276.56 & 161.48 & 1.20 \\ 
  \hline
 &BP2 & 474.20 & 240.09 & 450.80 & 239.40 & 229.60 & 167.60 & 1.15 \\
  \hline
  &BP3 & 490.39 & 365.26 & 483.68 & 310.39 & 328.30 & 272.23 & 1.07 \\
  \hline
 \end{tabular}
\caption{Masses for different non-standard scalars and the corresponding SM
  Higgs signal strength within 2$\sigma$ limit of the current value
  $1.02^{+0.18}_{-0.19}$~\cite{Sirunyan:2018ouh}.}
\label{table:benchmark}
\end{table}

{\magenta \begin{table}[ht]
\centering
 \begin{tabular}{|c|c|c|c|c|}
  \hline
&  & sin $\alpha$ &  sin $\beta$ &  sin $\gamma$ \\
  \hline
 & BP1 & 0.76 & 0.63 & 0.55  \\ 
  \hline
 &BP2 & 0.66 & 0.65 & 0.35   \\
  \hline
  &BP3 & 0.73 & 0.70 & 0.61   \\
  \hline
 \end{tabular}
\caption{Mixing angles between for the doubly charged ($\alpha$), 
singly charged ($\beta$) and CP-even(CP-odd) scalars($\gamma$).} \label{table:mixing angles}
\end{table}}

Even after fixing fixed all  model 
parameters in Table~\ref{table:fixedparam}, the Yukawa 
coupling matrices $h^{(1)}$ and $h^{(2)}$ still remain indeterminate (Eq.~\ref{ymat}).
We fix the matrix $h^{(2)}$ 
by choosing one suitable value for all elements of the 
$\mu$--$\tau$ block and keeping the rest of the elements one order smaller. 
It is emphasized that this {\it ad hoc} convention does not
affect the generality of our results. One may thus write
\begin{eqnarray*}
h^{(1)}_{ij} &=&
\left(\begin{array}{ccc}
4.52 \times 10^{-12} & 1.02 \times 10^{-11} & 3.47 \times 10^{-12} 
\\ 
1.02 \times 10^{- 11}  & 2.12 \times 10^{-11} & 1.90 \times 10^{-11}  
\\ 
3.47 \times 10^{- 12}  & 1.90 \times 10^{- 11} & 3.68 \times 10^{- 11}
\end{array}\right),
\\
h^{(2)}_{ij} &=& 
\left(\begin{array}{ccc}
1.0 \times 10^{-12} & 1.0 \times 10^{-12} & 1.0 \times 10^{-12} 
\\ 
1.0 \times 10^{- 12}  & 1.0 \times 10^{-11}  & 1.0 \times 10^{-11} 
\\ 
1.0 \times 10^{- 12}  & 1.0 \times 10^{- 11} & 1.0 \times 10^{- 11}
\end{array}\right).
\end{eqnarray*}

\subsection{Collider search at the LHC}
\label{sec:coll_lhc}
We finally turn to  
signals of this scenario at the high energy high luminosity (HE-HL) 
run of the LHC. There are several production and decay chains that lead to 
our chosen final state Eq.(~\ref{channel}) 
\begin{subequations}\footnote{ For benchmark points BP1 and BP3,
the main contribution comes from ~\ref{pair2} as
${\rm BR}(H_{1}^{++}\to H^+_2 W^+)\simeq 99\%$. However, situation is different for BP2, where, 
${\rm BR}(H^{++}_1 \rightarrow H^{++}_2 h) \simeq 23 \%$, making 
all the four production and decay modes significant.} 

{\bf
{\small
\begin{eqnarray}
p p \to H^{++}_1 H^{--}_1 &\to & (H^{++}_2 h) + (H^{--}_2 h) \to  (W^+ W^+ h) + (W^- W^- h) \to  2(\ell^{\pm} \ell^{\pm}) 
+ 4b + \mET \,\label{pair1}~~\\
p p \to H^{++}_1 H^{--}_1 &\to & (H^{+}_2 W^+) + (H^{-}_2 W^-) \to  (W^+ W^+ h) + (W^- W^- h) \to  2(\ell^{\pm} \ell^{\pm}) 
+ 4b + \mET~\label{pair2}~~~~\\
p p \to H^{++}_1 H^{--}_1 &\to & (H^{+}_2 W^+) + (H^{--}_2 h) \to  (W^+ W^+ h) + (W^- W^- h) \to  2(\ell^{\pm} \ell^{\pm}) 
+ 4b + \mET \,\label{mixed1}~~\\
p p \to H^{++}_1 H^{--}_1 &\to & (H^{++}_2 h) + (H^{-}_2 W^-) \to  (W^+ W^+ h) + (W^- W^- h) \to  2(\ell^{\pm} \ell^{\pm}) 
+ 4b + \mET \,\label{mixed2}.
\end{eqnarray}
}}
\end{subequations}
It should be noted that $Br(H_2^{+} \to W^{+} Z)$ and $Br(H_2^{+} \to t \bar{b})$ could give rise 
to the same final state. However, $Br(H_2^{+} \to W^{+} h)$
is 99.99$\%$, 97.60$\%$ and 99.5$\%$ for BP1, BP2 and BP3 respectively. The $W^{+} h$ mode dominates over $W^{+} Z$ and
$t \bar{b}$ as the $W^{+} h$ channel receives a single-fold suppression due to doublet-triplet mixings while the two remaining 
channels ($W^{+} Z$, $t \bar{b}$) have two-fold suppression, namely, that from mixing as well as by the small triplet vev which
explicitly enters into the couplings. There is a further cancellation between the two above mentioned effects for the $W^{+} Z$ and
$t \bar{b}$ channels making the branching fraction negligible for these two decay modes.

We calculate the event rates in
Madgraph5(v2.4.3)~\cite{Alwall:2014hca} with the 
appropriate Feynman rules obtained via
{\tt FeynRules} ~\cite{Alloul:2013bka}.  
The signal as well as all the relevant 
standard model background events are calculated at the lowest order (LO)
with {\tt CTEQ6L}~\cite{Pumplin:2002vw} parton distribution functions, setting the 
renormalization and factorization scales at $M_Z$.
They are subsequently multiplied by the next-to-leading order (NLO) K-factors 
for the signal and the SM background processes,
taken as 1.25 \cite{ATLAS:2016pbt} and 
1.3~\cite{Campbell:2012dh, Lazopoulos:2008de , Frixione:2015zaa} respectively. 
For the showering and hadronization of both the signal and the SM background 
events we use the {\tt Pythia}(v6.4)~\cite{Sjostrand:2006za}, and the
detector simulation is done in {\tt Delphes}(v3)~\cite{deFavereau:2013fsa}, where
jets are constructed using the anti-$K_{T}$  algorithm
~\cite{Cacciari:2008gp}. The cut-based analyses are done using the 
{\tt MadAnalysis5}~\cite{Conte:2012fm}.
The production of $t{\bar t} Z$,  $t{\bar t} W^{\pm}$ and $t{\bar t} h$ constitute
the dominant SM background for our signal.  
While generating events, 
we select jets and leptons (electron and muon) 
using the following kinematical acceptance cuts : 
\begin{subequations}
\begin{eqnarray}
\Delta R_{j j} > 0.6,~~~~\Delta R_{\ell \ell} > 0.4,~~~~\Delta R_{j \ell} > 0.7\,, \\
\Delta R_{bj} > 0.7,~~~~\Delta R_{b \ell} > 0.2,\\ 
p_{T_{\rm min}}^j > 20 ~{\rm GeV},~~~~|\eta_j| < 5,\\
p_{T_{\rm min}}^\ell > 10 ~{\rm GeV},~~~~|\eta_\ell| < 2.5,
\end{eqnarray}\label{basic_cuts}
\end{subequations}


The presence of four b-jets in the signal makes
b-tagging an important issue.
For this we
comply with the efficiency formula proposed by the ATLAS
collaboration~\cite{ATLAS:2014pla} for both the signal and background
processes as follows:
\begin{eqnarray}
\epsilon_b = \begin{cases}
0 & p_T^b \leq 30 ~\rm GeV \\
0.6 & 30 ~{\rm GeV} < p_T^b < 50 ~{\rm GeV} \\
0.75 & 50 ~{\rm GeV} < p_T^b < 400 ~{\rm GeV} \\
0.5 & p_T^b > 400 ~\rm GeV~.
\end{cases} 
\label{b-tag}
\end{eqnarray}

In addition, a mistagging probability of 10\% (1\%) for charm-jets
(light-quark and gluon jets) as a $b$-jet has been taken into account.
For lepton isolation, we abide by the criteria defined in
Ref.~\cite{ATLAS:2016rin} where the electrons are isolated with the
{\tt Tight} criterion defined in Ref.~\cite{ATLAS:2016iqc} and the
muons are isolated using the {\tt Medium} criterion defined in
reference~\cite{Aad:2016jkr}.

\begin{figure}[htbp!]
		\begin{tabular}{cc}
		\includegraphics[width=9cm,height=7cm]{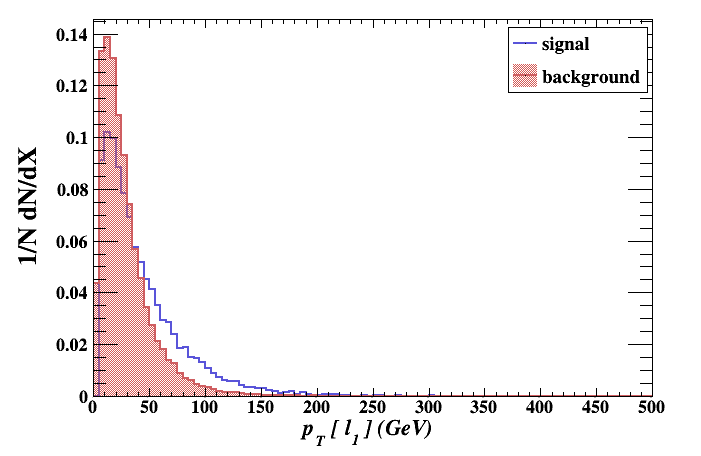} 
		\includegraphics[width=9cm,height=7cm]{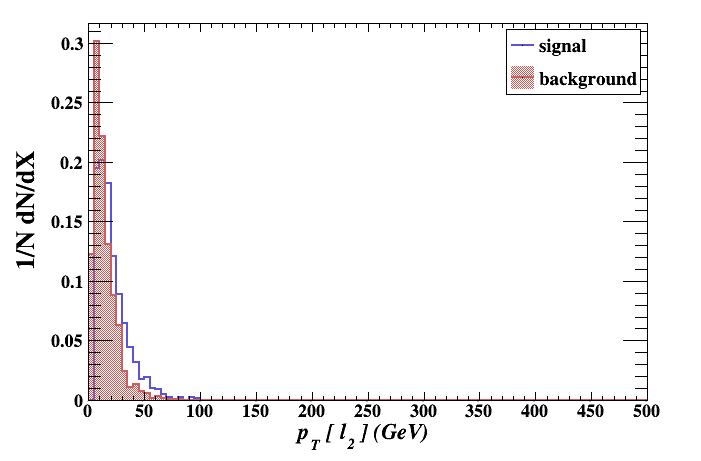}
		\end{tabular}
	\caption{\textit{Transverse momentum $(p_T)$ distribution (normalized) of the leading and subleading leptons
	for the benchmark BP1.}}
	\label{fig:lep} 
\end{figure}

One point worth mentioning at this point is
that we consider the all inclusive decay channels for both the signal
and background event generation, not only the leptonic decay of the SM $W^\pm$ boson. This is true for all the subsequent analyses. 
Before reporting the results in detail, let us examine  some kinematic distributions
relevant for the analysis,
starting with transverse momenta ($p_T$) of 
the two leading leptons as depicted in Figure~\ref {fig:lep} for  
BP1. The signal and background distributions are shown in blue and 
red respectively. In the signal events, these leptons originate
from the decay of the $W$ boson, while for the SM background processes,
they come from the decay of $W^\pm $ or  $Z $-bosons. From the shape
of the $p_T$ distribution one can see that it is evidently difficult to 
impose any selection cut on the $p_T$ of these two leading leptons to 
distinguish the signal from the SM backgrounds. This general feature 
is also found for other two benchmarks points. Hence, we 
only put the basic acceptance cut on the $p_T$ of the leptons, $p_T > 10 $ GeV. 

\begin{figure}[ht!]
\centering
\includegraphics[width=8cm,height=6cm, angle=0]{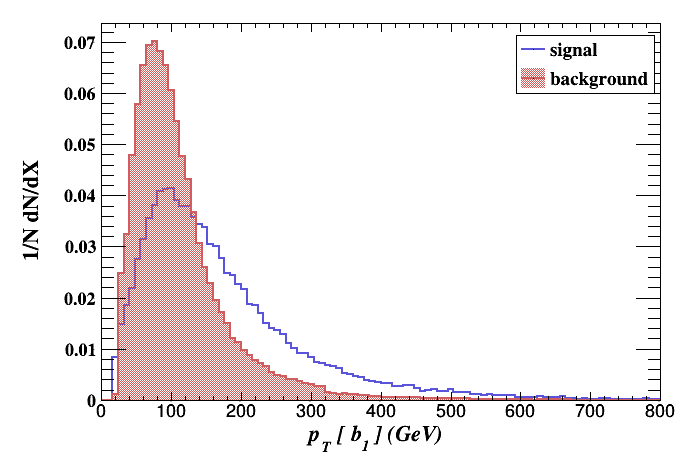} 
\includegraphics[width=8cm,height=6cm, angle=0]{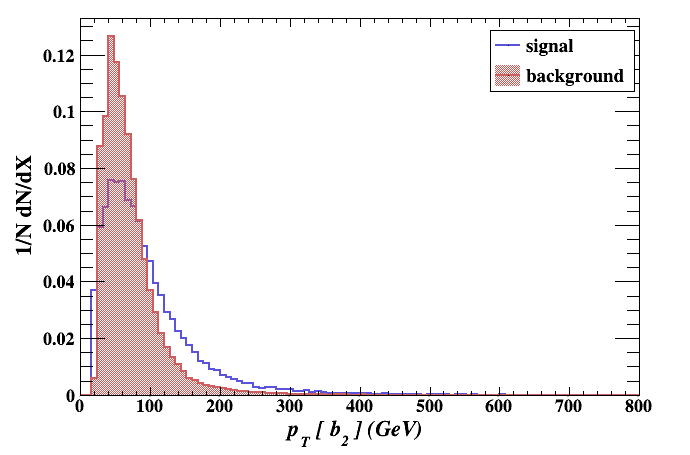} \\
\includegraphics[width=8cm,height=6cm, angle=0]{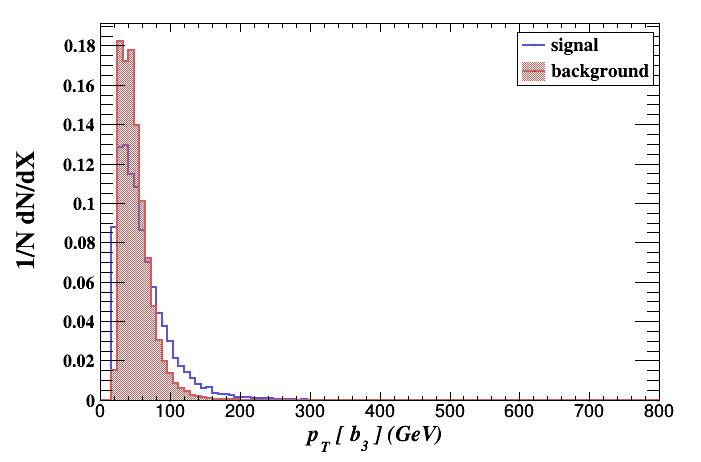} 
\includegraphics[width=8cm,height=6cm, angle=0]{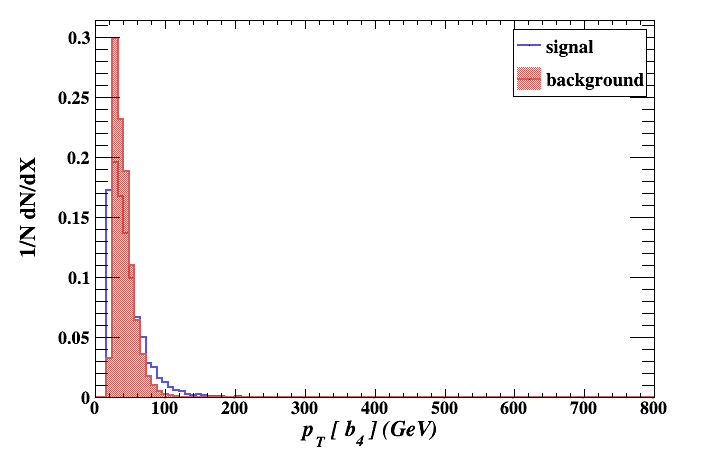}
 \caption{\label{fig:jet_pT} \textit{Transverse momentum $(p_T)$
     distribution (normalized) of the four leading $b~$jets for the
     final state $ 2\ell^\pm \ell^\pm+ 4 b + \mET $ for the benchmark
     BP1.  }}
 \end{figure}

In Figure\;\ref{fig:jet_pT} we show the $P_T$ distributions of four
$b-$jets for both signal and background events. In the signal events,
they come from decays of two SM-like scalars each of which is produced
via either from $H_1^{\pm\pm} \to H_2^{\pm\pm}h \to W^{\pm}W^{\pm}h$ or from 
$H_1^{\pm\pm} \to H_2^{\pm}W^{\pm} \to W^{\pm}W^{\pm}h$.  On the other hand,
the background $b$-jets have their sources mostly in $t{\bar t}h$,
$t{\bar t}W^{\pm}$ and $t{\bar t}Z$.  In the case of $t{\bar t} W^{\pm}$ production, two
$b$-jets come from the top quarks, while a pair of light quark jets may
fake as $b$-jets. The leading $b$-jet of our signal
events is found to be harder than that of those in the background. 
Thus we
demand the leading $b$-jet to have $p_T(j_1) > 80$ GeV, and
$p_T$ of the subsequent 3 $b$-jets to be
$p_T(j_2) > 60$ GeV, $p_T(j_3) > 30$ GeV and $p_T(j_4) > 20$ GeV.

\begin{figure}[ht!]
\centering
\includegraphics[width=8cm,height=7cm, angle=0]{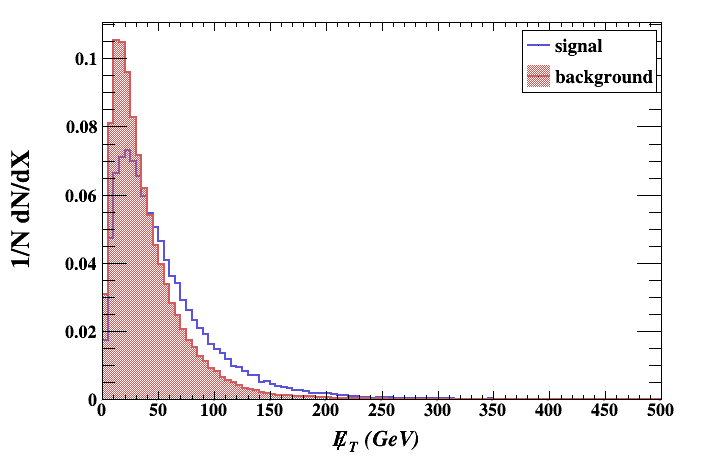} 
\includegraphics[width=8cm,height=7cm, angle=0]{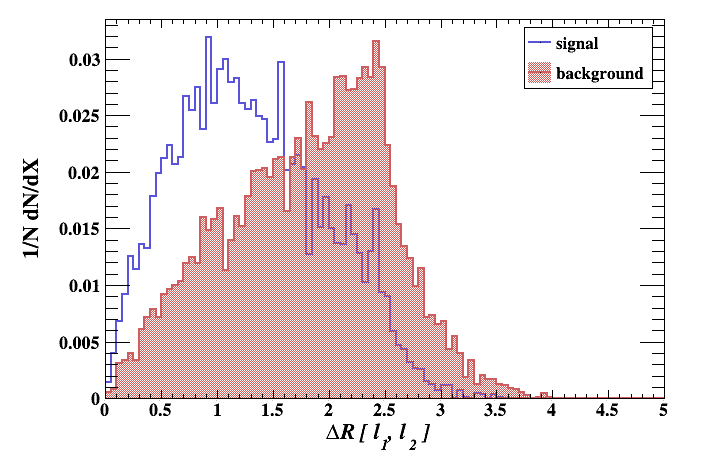}
\caption{\label{fig:dr} \textit{The normalized missing transverse energy distribution (left panel) $\mET$ and 
$\Delta R(\ell_1^\pm \ell_2^\pm)$ distribution (normalized) (right panel)between the two same-sign 
leptons for the benchmark BP1.
}}
 \end{figure}

The normalized distribution of the missing transverse energy 
($\mET$) are in the left panel of Figure~\ref{fig:dr}. One should note that the  
$\mET $ for both signal and backgrounds is due to either  
neutrinos or the missmeasurement of the jet and 
lepton momenta. Consequently, the 
shape of the distributions for both the signal and the background 
look very similar. The small rightward shift of the peak of 
the signal distribution can be attributed to fact that $W^{\pm}$ are boosted 
as they are produced from the decays of much heavier parent scalars. 
Hence, we find that a moderate requirement of $\mET > 30$ GeV is 
sufficient to improve the signal to background ratio.  
 
The right panel of Figure~\ref{fig:dr} shows the normalized distribution of 
angular separation $\Delta R(\ell^\pm\ell^\pm)$ between two same-sign 
leptons for both the signal and the background events. 
For our signal events, both leptons come from  
same-sign $W^\pm$ which are produced from the decay of heavy doubly charged
scalar with non-zero transverse momenta. The $P_T$ of the heavy doubly 
charged scalar forces its decay products ($W$-pair) to appear with 
relatively small $\Delta R$. The resultant effect of these two is 
finally displayed by the two same-sign leptons that also appear with 
small opening angle. For the SM 
background, on the other hand, they come from 
$W^\pm/Z$ bosons radiated from top/anti-top quarks, and can 
have wider separations. An upper cut on 
$\Delta R(\ell_1^\pm \ell_2^\pm )< 1.5$ thus enhances our 
signal-to-background ratio.

The cuts are summarised below, in the order in which they are 
imposed: 
 \begin{itemize}
  \item ({\bf C-1}): We want the leading $b$-jet to have $p_T(b_1) > 80$ GeV.
     This is motivated by the top-left panel of 
    Figure~\ref{fig:jet_pT} and it immediately reduces the SM background 
    arising from $t{\bar t} V, V \equiv W^\pm, Z$ processes by almost $50\%$. 

  \item ({\bf C-2}): Given the fact that the second leading $b$-jet for the
signal is not very hard, we demand that $p_T(b_2) > 60$ GeV. 
This cut also enhances the signal to background ratio to a reasonable extent. 
  
\item ({\bf C-3}): $p_T (b_3)$ is chosen to be $ > 30$ GeV. 
                   
\item ({\bf C-4}): It is evident from the bottom-right panel of 
Figure\;\ref{fig:jet_pT} that the fourth $b$ jet is very soft. So, 
the choice of $p_T(b_4) > 20$ GeV ensures the presence of four 
$b$ jets in the signal.
    
\item ({\bf C-5}): We select two same sign leptons with $p_T > 10$ GeV. 
     
\item ({\bf C-6}): We apply a veto on any oppositely charged lepton 
       with $p_T(\ell) > 10$ GeV.
      
 \item ({\bf C-7}): $\mET > 30$ GeV is imposed.
This also takes care of fake $\mET$.
       
\item ({\bf C-8}): The most effective cut to reduce the SM backgrounds 
is the angular separation between the same sign dileptons. 
The requirement of $\Delta R (\ell^\pm_1 \ell^\pm_2 ) < $ 1.5 considerably 
improves the signal to noise ratio.   
 \end{itemize}
 
With these cuts imposed, we  obtain the statistical significance of 
the signal HE-HL 
LHC with $\sqrt{s} = 14 $ TeV and also present some tentative predictions 
for the proposed FCC-hh with $ \sqrt{s} = 100 \rm~TeV$.
 
\subsubsection{Collider search at the LHC at $ \sqrt{s} = 14 \rm TeV$}

Table ~\ref{signi} contains the cut flow 
for $ \sqrt{s} = 14~\rm TeV$. The 
statistical significance is given by 
\begin{eqnarray}
{\cal S} = \sqrt{2 \times [(n_s+n_b) ~{\rm ln}(1+\frac{n_s}{n_b}) - n_s]},
\label{eq:signi}
\end{eqnarray}
where $n_s(n_b)$ denotes the number of signal (background)
events after implementing all the cuts at a specific luminosity.
The signal significance can be seen for three 
benchmark points,  assuming an integrated luminosity ${\cal L}_{\rm int}$ of 
$3~{\rm ab}^{-1}$. We obtain the highest statistical significance 
${\cal S} = 5.80$ for BP1, which corresponds to
$m_{H_1^{\pm\pm}} = 416.37 $ GeV and $m_{H_2^{\pm\pm}} = 216.13 $ GeV
respectively. For the other two benchmark points with relatively heavier
scalar masses, the significances are $3.03\sigma$ and $1.84\sigma$ respectively, which may   
be treated as the evidence of the two-triplet scenario. 
\begin{table}[htbp!]
	\centering
	\footnotesize
\resizebox{17cm}{!}{
	\begin{tabular}{|p{1.7cm}|p{2.1cm}|c|c|c|c|c|c|c|c|p{1.7cm}|}
		\cline{1-10}
		\multicolumn{2}{|c|}{}& \multicolumn{8}{|c|}{Effective cross section (fb) after the cut}   \\ \cline{1-10}
		SM-background & Production Cross section (fb) 
		& C1--1 & C1--2 & C1--3 & C1--4 & C1--5 & C1--6 & C1--7 & C1--8 
		  \\ \cline{1-10}
		
		$t\bar{t}+W^\pm$ & 517.4 & 280.59 & 96.81 & 11.27 & 0.51 & 0.0028 & 0.0023 & 0.002 & 0.0001 \\ \cline{1-10}
		$t\bar{t}+Z$ & 917.8 & 436.06 & 214.72 & 56.28 & 10.37 & 0.009 & 0.0052   & 0.004 & 0.0004 \\ \cline{1-10}
		$t\bar{t}+h$ & 622.7 & 535.21 & 242.88 & 121.93 & 31.36 & 0.0026 & 0.0016    & 0.001 & 0.0001 \\  \cline{1-10}
		\multicolumn{1}{|c|}{Total SM Background} & 2058 & 1251.86 & 554.41 & 189.48 & 42.24 & 0.0144 & 0.0091  & 0.007 & 0.0006   \\ \cline{1-9}  \hline
	\multicolumn{1}{|c|}{Signal}&{Production Cross section (fb)}& \multicolumn{8}{|c|}{Effective cross section (fb) for signal after the cut} & {Significance reach at ${\cal L}_{\rm int}=3~{\rm ab}^{-1}$}\\ \hline
	\multicolumn{1}{|c|}{BP1} & 4.41 & 3.59 & 2.44 & 1.56 & 0.66 & 0.011 & 0.009 & 0.008 & 0.0042 & 5.80 \\ \hline 
	\multicolumn{1}{|c|}{BP2} & 2.08 & 1.71 & 1.16 & 0.74 & 0.32 & 0.0051 & 0.0044 & 0.0035 & 0.0018 & 3.03 \\ \hline 
	\multicolumn{1}{|c|}{BP3} & 1.05 & 0.86 & 0.58 & 0.37 & 0.16 & 0.0025 & 0.0019 & 0.0016 & 0.0010 & 1.84 \\ \hline \hline 
	\end{tabular}}
	\caption{\it Effective cross section obtained after each cut for both signal ( $2(\ell^{\pm} \ell^{\pm}) 
+ 4b + \mET$) and background and the 
   	respective significance reach at $3~{\rm ab}^{-1}$ integrated luminosity at $\rm 14~TeV$ LHC}
	\label{signi}
\end{table}

\begin{figure}[ht!]
\centering
\includegraphics[width=10cm,height=7cm, angle=0]{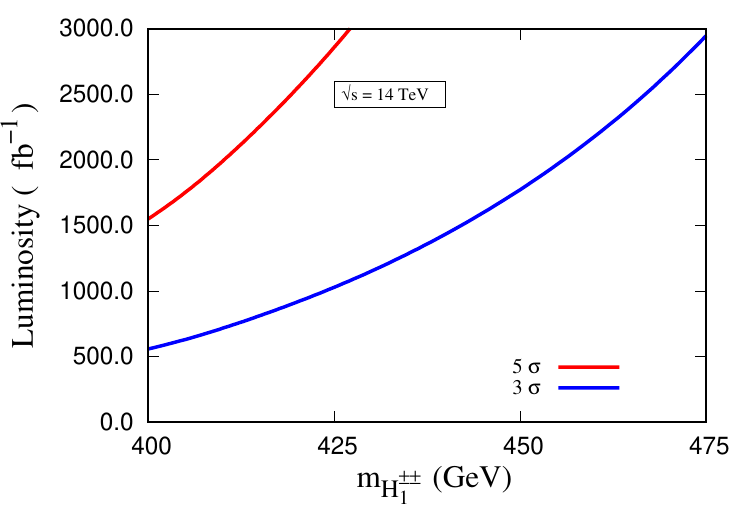} 
\caption{\label{fig:lumi} \textit{ Variation of the  
integrated luminosity $({\cal L}_{\rm int})$ as a function of the 
doubly charged scalar mass($m_{H_1^{\pm\pm}}$) at the 14 TeV LHC.
The red and blue lines correspond to the ${\cal L}_{\rm int}$ required 
for the $5 \sigma$ discovery and $3 \sigma$ evidence respectively for 
the signal studied.}}
 \end{figure}
 
In Figure~\ref{fig:lumi} we present the doubly charged scalar($H_1^{\pm\pm}$) 
mass and the integrated luminosity(${\cal L}_{\rm int}$) required to reach 
$5 \sigma$(red solid line) and $3 \sigma$( blue solid line)  
significance with 14 TeV. It is evident from this 
Figure that  `discovery' is not possible beyond $m_{H^{\pm\pm}_1}  \sim 430$ GeV even with $3 \rm ab^{-1}$. 
However, one can probe the doubly charged scalar mass up to $475 $~GeV
at $3 \sigma$ level.

\begin{figure}[ht!]
\centering
\includegraphics[width=12cm,height=8cm, angle=0]{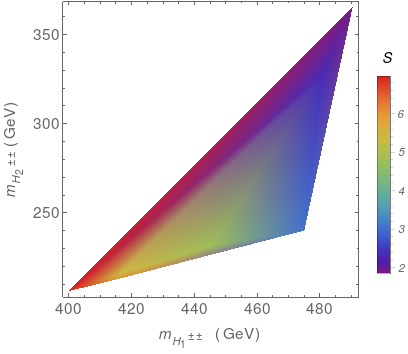} 
\caption{\label{fig:contour} \textit{
Signal significance $({\cal S})$ in the 
$m_{H_1^{\pm\pm} }- m_{H_2^{\pm\pm}} $ plane, shown with continuously 
varying colour codes, for $3~{\rm ab}^{-1}$ integrated luminosity at 14 TeV.}}
 \end{figure}

Figure~\ref{fig:contour} shows the variation of
${\cal S }$ in the $m_{H_1^{\pm\pm} }- m_{H_2^{\pm\pm}} $ plane. 
${\cal S}$ drops from $7 \sigma$ to $2 \sigma$ as mass of 
$H_1^{\pm\pm}(H_2^{\pm\pm}$) is increased from $400(206)$ GeV to 
$490(370)$ GeV due to  phase space suppression.
It is worth mentioning here that, for the single-triplet 
case with $w \sim 1 $ GeV, doubly charged scalar 
masses up to 300 GeV can be explored at the $5\sigma$ level at 
the LHC with an integrated luminosity of 
$3~{\rm ab}^{-1}$~\cite{Ghosh:2017pxl}.

\newpage
\subsubsection{Search at $ \sqrt{s} = 100 $ TeV FCC-hh collider}

\begin{figure}[ht!]
\centering
\includegraphics[width=10cm,height=7cm, angle=0]{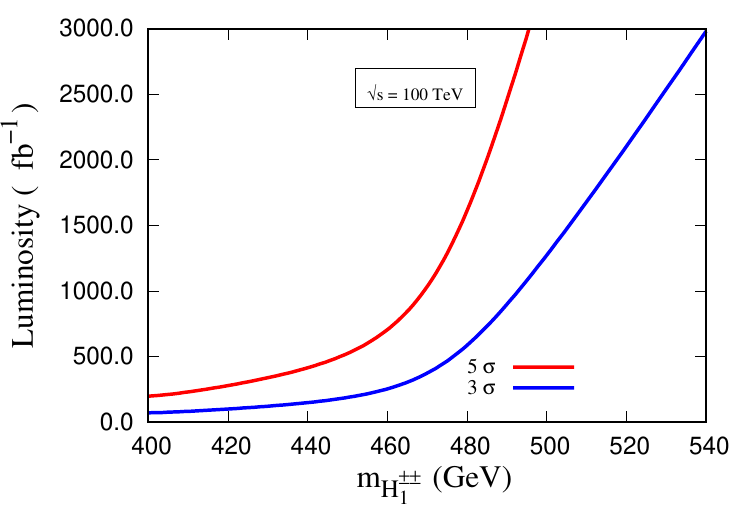} 
\caption{\label{fig:lumi_100} \textit{Same as in Figure \ref{fig:lumi} for the 100 TeV FCC-hh.}}
\end{figure}

We next make a tentative estimate of the potential of the proposed 100
TeV FCC-hh to investigate the two triplet scalar scenario. In our
numerical analysis, we apply the set of selection cuts as those used
for the 14 TeV LHC. Since the production mode of our signal processes are
electroweak processes, the NLO K-factors, too, are taken to be the
same, an assumption justifiable in the light of
\cite{Mangano:2016jyj}. Table
\ref{signi_100} summarises the effect of different selection cuts on
the signal and the SM backgrounds for all the three benchmark
points. At this energy, all of them can
be probed with ${\cal S} \geq 5 \sigma$ at $3{\rm ab}^{-1}$.  
A comparison with the 14 TeV LHC shows that, while a 
$5\sigma$ reach with  $( 3{\rm ab}^{-1})$ is
possible only for BP1, a distinct improvement is foreseen for
the FCC-hh. This is reflected in Figure~\ref{fig:lumi_100} 
which shows that one can probe doubly charged Higgs masses up 
to  $495(540)$ GeV at the $5 \sigma (3 \sigma)$ significance level.

\begin{table}[htbp!]
	\centering
	\footnotesize
\resizebox{17cm}{!}{
	\begin{tabular}{|p{1.7cm}|p{2.1cm}|c|c|c|c|c|c|c|c|p{1.7cm}|}
		 \cline{1-10}
		\multicolumn{2}{|c|}{}& \multicolumn{8}{|c|}{Effective cross section (fb) after the cut}   \\ \cline{1-10}
		SM-background & Production Cross section (fb) 
		& C2--1 & C2--2 & C2--3 & C2--4 & C2--5 & C2--6 & C2--7 & C2--8 
		\\ \cline{1-10}
		$t\bar{t}+W^\pm$ & $2.07\times 10^4$ &  $6.27\times 10^3$ &  $1.18\times 10^3$ &  $1.84\times 10^2$ & 10.73 & 0.10 & 0.081 & 0.040 & 0.004 \\ \cline{1-10}
		$t\bar{t}+Z$ &  $6.41\times 10^4$ & $3.10\times 10^4$ & $1.12\times 10^4$ & $2.77\times 10^3$ & 55.09 & 0.13 & 0.12   & 0.116 & 0.009 \\ \cline{1-10}
		$t\bar{t}+h$ & $3.80\times 10^4$ & $2.46\times 10^4$ & $1.24\times 10^4$ & $5.68\times 10^3$ & $1.51\times 10^3$ & 0.23 & 0.22  & 0.21 & 0.015 \\  \cline{1-10}
		\multicolumn{1}{|c|}{Total SM Background} & $1.11\times 10^5$ & $6.19\times 10^4$ & $2.48\times 10^4$ & $8.63\times 10^3$ & $1.57\times 10^3$ &0.46 & 0.42  & 0.366 & 0.028  \\ \cline{1-9}  \hline
	\multicolumn{1}{|c|}{Signal}&{Production Cross section (fb)}& \multicolumn{8}{|c|}{Effective cross section (fb) for signal after the cut} & 
	{Significance reach at ${\cal L}_{\rm int}=3~{\rm ab}^{-1}$}\\ \hline
	\multicolumn{1}{|c|}{BP1} & 125.0 & 83.59 & 51.08 & 30.29 & 12.35 & 0.142 & 0.122 & 0.121 & 0.072 &  18.21 \\ \hline 
	\multicolumn{1}{|c|}{BP2} & 68.08 & 45.90 & 27.99 & 16.60 & 6.67 & 0.077 & 0.066 & 0.065 & 0.039 & 10.80 \\ \hline 
	\multicolumn{1}{|c|}{BP3} & 28.81 & 19.35 & 11.83 & 7.02 & 2.88 & 0.034 & 0.029 & 0.028 & 0.020 & 5.94 \\ \hline \hline 
	\end{tabular}}
	\caption{\it Effective cross section obtained after each cut for both signal ( $2(\ell^{\pm} \ell^{\pm}) 
+ 4b + \mET$) and background and the 
   	respective significance reach at $3~{\rm ab}^{-1}$ integrated luminosity at $\rm 100~TeV$ pp collider}
	\label{signi_100}
\end{table}

\begin{figure}[ht!]
\centering
\includegraphics[width=12cm,height=8cm, angle=0]{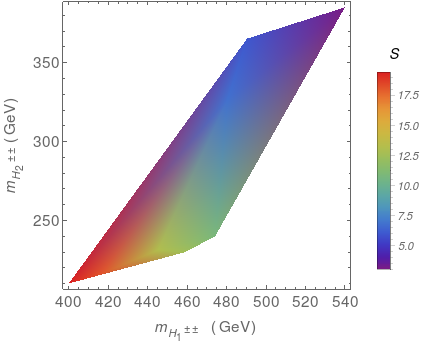} 
\caption{\label{fig:contour_100} \textit{Same as in Figure \ref{fig:contour} for the 100 TeV FCC-hh.}}
 \end{figure}

The signal topology 
studied in this work strongly depends on $m_{H^{\pm \pm}_1}$, 
$m_{H^{\pm \pm}_2}$ and their mass splitting. This is accentuated 
in Figure~\ref{fig:contour_100} where we show the dependence 
of the signal significance on these two masses. using colour-codes
significance regions once more. 

\section{Conclusion}
We have extended the Type-II seesaw model with an extra $SU(2)_L$
complex triplet scalar and allow a small mixing of the triplets 
with the SM Higgs doublet as required by electroweak precision
measurements. After EWSB, one has a rather rich scalar spectrum
including two each of doubly and singly charged scalars. A collider signal 
distinctly reflecting the presence of two triplets is identified in the form of
$2\ell^\pm \ell^\pm+ 4 b + \mET$.  This
channel can arise both from $H_1^{\pm\pm} \rightarrow H_2^{\pm\pm} h$
and $H_1^{\pm\pm} \rightarrow H_2^{\pm} W^{\pm}$ following  the
characteristic decay modes in the case where the Yukawa couplings are too small (large triplet vev)
to trigger the dominant same-sign-dilepton channels.
Three benchmark points, consistent with all phenomenological constraints,
have been used as illustration. We have estimated the potential at high luminosity 
LHC and a high energy pp collider $100$ TeV in identifying this signal.

We conclude after a cut-based analysis that the $5\sigma $ discovery
reach is possible with 3 ${\rm ab}^{-1}$ luminosity at $14$ TeV LHC
for the mass of the heavier doubly charged scalar up to about 430
GeV. For the proposed FCC-hh collider operating at $\sqrt{s} = 100$
TeV, this reach can be extended up to 495 GeV. Though we have not
taken into account some experimental issues such as jet faking
leptons, lepton charge misidentification and photon conversions into
lepton pairs, these effects are unlikely to affect the predictions
qualitatively.  Therefore, with appropriate refinement, our prediction
should help in probing this scenario which is relevant in framing
predictive models for neutrino masses and mixing.


\section{Acknowledgement}
NG would like to acknowledge the Council of Scientific and Industrial
Research (CSIR), Government of India for financial support. BM
acknowledges financial support from the Department of Atomic Energy,
Government of India, for the Regional Centre for Accelerator-based
Particle Physics (RECAPP), Harish-Chandra Research Institute. DKG and
NG thank RECAPP for hospitality while this work was being carried
out. BM would like to thank Indian Association for he Cultivation of
Science for hospitality during the concluding part of the project.


\begin{thebibliography}{10}

\bibitem{Aad:2012tfa}
{\scshape ATLAS} collaboration, G.~Aad et~al., \emph{{Observation of a new
  particle in the search for the Standard Model Higgs boson with the ATLAS
  detector at the LHC}},
  \href{https://doi.org/10.1016/j.physletb.2012.08.020}{\emph{Phys. Lett.}
  {\bfseries B716} (2012) 1--29},
  [\href{https://arxiv.org/abs/1207.7214}{{\ttfamily 1207.7214}}].

\bibitem{Chatrchyan:2012xdj}
{\scshape CMS} collaboration, S.~Chatrchyan et~al., \emph{{Observation of a new
  boson at a mass of 125 GeV with the CMS experiment at the LHC}},
  \href{https://doi.org/10.1016/j.physletb.2012.08.021}{\emph{Phys. Lett.}
  {\bfseries B716} (2012) 30--61},
  [\href{https://arxiv.org/abs/1207.7235}{{\ttfamily 1207.7235}}].

\bibitem{Walter:2008ys}
C.~W. Walter, \emph{{The Super-Kamiokande Experiment}},
  \href{https://arxiv.org/abs/0802.1041}{{\ttfamily 0802.1041}}.

\bibitem{GonzalezGarcia:2007ib}
M.~C. Gonzalez-Garcia and M.~Maltoni, \emph{{Phenomenology with Massive
  Neutrinos}}, \href{https://doi.org/10.1016/j.physrep.2007.12.004}{\emph{Phys.
  Rept.} {\bfseries 460} (2008) 1--129},
  [\href{https://arxiv.org/abs/0704.1800}{{\ttfamily 0704.1800}}].

\bibitem{Mohapatra:2006gs}
R.~N. Mohapatra and A.~Y. Smirnov, \emph{{Neutrino Mass and New Physics}},
  \href{https://doi.org/10.1146/annurev.nucl.56.080805.140534}{\emph{Ann. Rev.
  Nucl. Part. Sci.} {\bfseries 56} (2006) 569--628},
  [\href{https://arxiv.org/abs/hep-ph/0603118}{{\ttfamily hep-ph/0603118}}].

\bibitem{Minkowski:1977sc}
P.~Minkowski, \emph{{$\mu \to e\gamma$ at a Rate of One Out of $10^{9}$ Muon
  Decays?}}, \href{https://doi.org/10.1016/0370-2693(77)90435-X}{\emph{Phys.
  Lett.} {\bfseries 67B} (1977) 421--428}.

\bibitem{Yanagida:1979as}
T.~Yanagida, \emph{{HORIZONTAL SYMMETRY AND MASSES OF NEUTRINOS}}, {\emph{Conf.
  Proc.} {\bfseries C7902131} (1979) 95--99}.

\bibitem{Mohapatra:1979ia}
R.~N. Mohapatra and G.~Senjanovic, \emph{{Neutrino Mass and Spontaneous Parity
  Violation}}, \href{https://doi.org/10.1103/PhysRevLett.44.912}{\emph{Phys.
  Rev. Lett.} {\bfseries 44} (1980) 912}.

\bibitem{Schechter:1980gr}
J.~Schechter and J.~W.~F. Valle, \emph{{Neutrino Masses in SU(2) x U(1)
  Theories}}, \href{https://doi.org/10.1103/PhysRevD.22.2227}{\emph{Phys. Rev.}
  {\bfseries D22} (1980) 2227}.

\bibitem{Magg:1980ut}
M.~Magg and C.~Wetterich, \emph{{Neutrino Mass Problem and Gauge Hierarchy}},
  \href{https://doi.org/10.1016/0370-2693(80)90825-4}{\emph{Phys. Lett.}
  {\bfseries 94B} (1980) 61--64}.

\bibitem{Cheng:1980qt}
T.~P. Cheng and L.-F. Li, \emph{{Neutrino Masses, Mixings and Oscillations in
  SU(2) x U(1) Models of Electroweak Interactions}},
  \href{https://doi.org/10.1103/PhysRevD.22.2860}{\emph{Phys. Rev.} {\bfseries
  D22} (1980) 2860}.

\bibitem{Lazarides:1980nt}
G.~Lazarides, Q.~Shafi and C.~Wetterich, \emph{{Proton Lifetime and Fermion
  Masses in an SO(10) Model}},
  \href{https://doi.org/10.1016/0550-3213(81)90354-0}{\emph{Nucl. Phys.}
  {\bfseries B181} (1981) 287--300}.

\bibitem{Mohapatra:1980yp}
R.~N. Mohapatra and G.~Senjanovic, \emph{{Neutrino Masses and Mixings in Gauge
  Models with Spontaneous Parity Violation}},
  \href{https://doi.org/10.1103/PhysRevD.23.165}{\emph{Phys. Rev.} {\bfseries
  D23} (1981) 165}.

\bibitem{Foot:1988aq}
R.~Foot, H.~Lew, X.~G. He and G.~C. Joshi, \emph{{Seesaw Neutrino Masses
  Induced by a Triplet of Leptons}},
  \href{https://doi.org/10.1007/BF01415558}{\emph{Z. Phys.} {\bfseries C44}
  (1989) 441}.

\bibitem{Franceschini:2008pz}
R.~Franceschini, T.~Hambye and A.~Strumia, \emph{{Type-III see-saw at LHC}},
  \href{https://doi.org/10.1103/PhysRevD.78.033002}{\emph{Phys. Rev.}
  {\bfseries D78} (2008) 033002},
  [\href{https://arxiv.org/abs/0805.1613}{{\ttfamily 0805.1613}}].

\bibitem{delAguila:2008cj}
F.~del Aguila and J.~A. Aguilar-Saavedra, \emph{{Distinguishing seesaw models
  at LHC with multi-lepton signals}},
  \href{https://doi.org/10.1016/j.nuclphysb.2008.12.029}{\emph{Nucl. Phys.}
  {\bfseries B813} (2009) 22--90},
  [\href{https://arxiv.org/abs/0808.2468}{{\ttfamily 0808.2468}}].

\bibitem{Arhrib:2009mz}
A.~Arhrib, B.~Bajc, D.~K. Ghosh, T.~Han, G.-Y. Huang, I.~Puljak et~al.,
  \emph{{Collider Signatures for Heavy Lepton Triplet in Type I+III Seesaw}},
  \href{https://doi.org/10.1103/PhysRevD.82.053004}{\emph{Phys. Rev.}
  {\bfseries D82} (2010) 053004},
  [\href{https://arxiv.org/abs/0904.2390}{{\ttfamily 0904.2390}}].

\bibitem{Akeroyd:2005gt}
A.~G. Akeroyd and M.~Aoki, \emph{{Single and pair production of doubly charged
  Higgs bosons at hadron colliders}},
  \href{https://doi.org/10.1103/PhysRevD.72.035011}{\emph{Phys. Rev.}
  {\bfseries D72} (2005) 035011},
  [\href{https://arxiv.org/abs/hep-ph/0506176}{{\ttfamily hep-ph/0506176}}].

\bibitem{Akeroyd:2009hb}
A.~G. Akeroyd and C.-W. Chiang, \emph{{Doubly charged Higgs bosons and
  three-lepton signatures in the Higgs Triplet Model}},
  \href{https://doi.org/10.1103/PhysRevD.80.113010}{\emph{Phys. Rev.}
  {\bfseries D80} (2009) 113010},
  [\href{https://arxiv.org/abs/0909.4419}{{\ttfamily 0909.4419}}].

\bibitem{Akeroyd:2010je}
A.~G. Akeroyd and C.-W. Chiang, \emph{{Phenomenology of Large Mixing for the
  CP-even Neutral Scalars of the Higgs Triplet Model}},
  \href{https://doi.org/10.1103/PhysRevD.81.115007}{\emph{Phys. Rev.}
  {\bfseries D81} (2010) 115007},
  [\href{https://arxiv.org/abs/1003.3724}{{\ttfamily 1003.3724}}].

\bibitem{Akeroyd:2011zza}
A.~G. Akeroyd and H.~Sugiyama, \emph{{Production of doubly charged scalars from
  the decay of singly charged scalars in the Higgs Triplet Model}},
  \href{https://doi.org/10.1103/PhysRevD.84.035010}{\emph{Phys. Rev.}
  {\bfseries D84} (2011) 035010},
  [\href{https://arxiv.org/abs/1105.2209}{{\ttfamily 1105.2209}}].

\bibitem{Lindner:2016bgg}
M.~Lindner, M.~Platscher and F.~S. Queiroz, \emph{{A Call for New Physics : The
  Muon Anomalous Magnetic Moment and Lepton Flavor Violation}},
  \href{https://doi.org/10.1016/j.physrep.2017.12.001}{\emph{Phys. Rept.}
  {\bfseries 731} (2018) 1--82},
  [\href{https://arxiv.org/abs/1610.06587}{{\ttfamily 1610.06587}}].

\bibitem{Arhrib:2011uy}
A.~Arhrib, R.~Benbrik, M.~Chabab, G.~Moultaka, M.~C. Peyranere, L.~Rahili
  et~al., \emph{{The Higgs Potential in the Type II Seesaw Model}},
  \href{https://doi.org/10.1103/PhysRevD.84.095005}{\emph{Phys. Rev.}
  {\bfseries D84} (2011) 095005},
  [\href{https://arxiv.org/abs/1105.1925}{{\ttfamily 1105.1925}}].

\bibitem{Aoki:2012jj}
M.~Aoki, S.~Kanemura, M.~Kikuchi and K.~Yagyu, \emph{{Radiative corrections to
  the Higgs boson couplings in the triplet model}},
  \href{https://doi.org/10.1103/PhysRevD.87.015012}{\emph{Phys. Rev.}
  {\bfseries D87} (2013) 015012},
  [\href{https://arxiv.org/abs/1211.6029}{{\ttfamily 1211.6029}}].

\bibitem{Perez:2008ha}
P.~Fileviez~Perez, T.~Han, G.-y. Huang, T.~Li and K.~Wang, \emph{{Neutrino
  Masses and the CERN LHC: Testing Type II Seesaw}},
  \href{https://doi.org/10.1103/PhysRevD.78.015018}{\emph{Phys. Rev.}
  {\bfseries D78} (2008) 015018},
  [\href{https://arxiv.org/abs/0805.3536}{{\ttfamily 0805.3536}}].

\bibitem{Melfo:2011nx}
A.~Melfo, M.~Nemevsek, F.~Nesti, G.~Senjanovic and Y.~Zhang, \emph{{Type II
  Seesaw at LHC: The Roadmap}},
  \href{https://doi.org/10.1103/PhysRevD.85.055018}{\emph{Phys. Rev.}
  {\bfseries D85} (2012) 055018},
  [\href{https://arxiv.org/abs/1108.4416}{{\ttfamily 1108.4416}}].

\bibitem{Aoki:2011pz}
M.~Aoki, S.~Kanemura and K.~Yagyu, \emph{{Testing the Higgs triplet model with
  the mass difference at the LHC}},
  \href{https://doi.org/10.1103/PhysRevD.85.055007}{\emph{Phys. Rev.}
  {\bfseries D85} (2012) 055007},
  [\href{https://arxiv.org/abs/1110.4625}{{\ttfamily 1110.4625}}].

\bibitem{ATLAS:2017iqw}
{\scshape ATLAS} collaboration, T.~A. collaboration, \emph{{Search for
  doubly-charged Higgs boson production in multi-lepton final states with the
  ATLAS detector using proton-proton collisions at
  $\sqrt{s}=13\,\mathrm{TeV}$}}, ATLAS-CONF-2017-053.

\bibitem{Han:2007bk}
T.~Han, B.~Mukhopadhyaya, Z.~Si and K.~Wang, \emph{{Pair production of
  doubly-charged scalars: Neutrino mass constraints and signals at the LHC}},
  \href{https://doi.org/10.1103/PhysRevD.76.075013}{\emph{Phys. Rev.}
  {\bfseries D76} (2007) 075013},
  [\href{https://arxiv.org/abs/0706.0441}{{\ttfamily 0706.0441}}].

\bibitem{Chakrabarti:1998qy}
S.~Chakrabarti, D.~Choudhury, R.~M. Godbole and B.~Mukhopadhyaya,
  \emph{{Observing doubly charged Higgs bosons in photon-photon collisions}},
  \href{https://doi.org/10.1016/S0370-2693(98)00743-6}{\emph{Phys. Lett.}
  {\bfseries B434} (1998) 347--353},
  [\href{https://arxiv.org/abs/hep-ph/9804297}{{\ttfamily hep-ph/9804297}}].

\bibitem{Cheung:2002gd}
K.~Cheung and D.~K. Ghosh, \emph{{Triplet Higgs boson at hadron colliders}},
  \href{https://doi.org/10.1088/1126-6708/2002/11/048}{\emph{JHEP} {\bfseries
  11} (2002) 048}, [\href{https://arxiv.org/abs/hep-ph/0208254}{{\ttfamily
  hep-ph/0208254}}].

\bibitem{Chiang:2012dk}
C.-W. Chiang, T.~Nomura and K.~Tsumura, \emph{{Search for doubly charged Higgs
  bosons using the same-sign diboson mode at the LHC}},
  \href{https://doi.org/10.1103/PhysRevD.85.095023}{\emph{Phys. Rev.}
  {\bfseries D85} (2012) 095023},
  [\href{https://arxiv.org/abs/1202.2014}{{\ttfamily 1202.2014}}].

\bibitem{Kanemura:2013vxa}
S.~Kanemura, K.~Yagyu and H.~Yokoya, \emph{{First constraint on the mass of
  doubly-charged Higgs bosons in the same-sign diboson decay scenario at the
  LHC}}, \href{https://doi.org/10.1016/j.physletb.2013.08.054}{\emph{Phys.
  Lett.} {\bfseries B726} (2013) 316--319},
  [\href{https://arxiv.org/abs/1305.2383}{{\ttfamily 1305.2383}}].

\bibitem{kang:2014jia}
Z.~Kang, J.~Li, T.~Li, Y.~Liu and G.-Z. Ning, \emph{{Light Doubly Charged Higgs
  Boson via the $WW^*$ Channel at LHC}},
  \href{https://doi.org/10.1140/epjc/s10052-015-3774-1}{\emph{Eur. Phys. J.}
  {\bfseries C75} (2015) 574},
  [\href{https://arxiv.org/abs/1404.5207}{{\ttfamily 1404.5207}}].

\bibitem{Kanemura:2014goa}
S.~Kanemura, M.~Kikuchi, K.~Yagyu and H.~Yokoya, \emph{{Bounds on the mass of
  doubly-charged Higgs bosons in the same-sign diboson decay scenario}},
  \href{https://doi.org/10.1103/PhysRevD.90.115018}{\emph{Phys. Rev.}
  {\bfseries D90} (2014) 115018},
  [\href{https://arxiv.org/abs/1407.6547}{{\ttfamily 1407.6547}}].

\bibitem{Kanemura:2014ipa}
S.~Kanemura, M.~Kikuchi, H.~Yokoya and K.~Yagyu, \emph{{LHC Run-I constraint on
  the mass of doubly charged Higgs bosons in the same-sign diboson decay
  scenario}}, \href{https://doi.org/10.1093/ptep/ptv071}{\emph{PTEP} {\bfseries
  2015} (2015) 051B02}, [\href{https://arxiv.org/abs/1412.7603}{{\ttfamily
  1412.7603}}].

\bibitem{Chen:2014qda}
C.-H. Chen and T.~Nomura, \emph{{Search for $\delta^{\pm\pm}$ with new decay
  patterns at the LHC}},
  \href{https://doi.org/10.1103/PhysRevD.91.035023}{\emph{Phys. Rev.}
  {\bfseries D91} (2015) 035023},
  [\href{https://arxiv.org/abs/1411.6412}{{\ttfamily 1411.6412}}].

\bibitem{Han:2015hba}
Z.-L. Han, R.~Ding and Y.~Liao, \emph{{LHC Phenomenology of Type II Seesaw:
  Nondegenerate Case}},
  \href{https://doi.org/10.1103/PhysRevD.91.093006}{\emph{Phys. Rev.}
  {\bfseries D91} (2015) 093006},
  [\href{https://arxiv.org/abs/1502.05242}{{\ttfamily 1502.05242}}].

\bibitem{Han:2015sca}
Z.-L. Han, R.~Ding and Y.~Liao, \emph{{LHC phenomenology of the type II seesaw
  mechanism: Observability of neutral scalars in the nondegenerate case}},
  \href{https://doi.org/10.1103/PhysRevD.92.033014}{\emph{Phys. Rev.}
  {\bfseries D92} (2015) 033014},
  [\href{https://arxiv.org/abs/1506.08996}{{\ttfamily 1506.08996}}].

\bibitem{Mitra:2016wpr}
M.~Mitra, S.~Niyogi and M.~Spannowsky, \emph{{Type-II Seesaw Model and
  Multilepton Signatures at Hadron Colliders}},
  \href{https://doi.org/10.1103/PhysRevD.95.035042}{\emph{Phys. Rev.}
  {\bfseries D95} (2017) 035042},
  [\href{https://arxiv.org/abs/1611.09594}{{\ttfamily 1611.09594}}].

\bibitem{Ghosh:2017pxl}
D.~K. Ghosh, N.~Ghosh, I.~Saha and A.~Shaw, \emph{{Revisiting the high-scale
  validity of the type II seesaw model with novel LHC signature}},
  \href{https://doi.org/10.1103/PhysRevD.97.115022}{\emph{Phys. Rev.}
  {\bfseries D97} (2018) 115022},
  [\href{https://arxiv.org/abs/1711.06062}{{\ttfamily 1711.06062}}].

\bibitem{Dev:2018kpa}
P.~S.~B. Dev and Y.~Zhang, \emph{{Displaced vertex signatures of doubly charged
  scalars in the type-II seesaw and its left-right extensions}},
  \href{https://arxiv.org/abs/1808.00943}{{\ttfamily 1808.00943}}.

\bibitem{Grimus:2004hf}
W.~Grimus, A.~S. Joshipura, L.~Lavoura and M.~Tanimoto, \emph{{Symmetry
  realization of texture zeros}},
  \href{https://doi.org/10.1140/epjc/s2004-01896-y}{\emph{Eur. Phys. J.}
  {\bfseries C36} (2004) 227--232},
  [\href{https://arxiv.org/abs/hep-ph/0405016}{{\ttfamily hep-ph/0405016}}].

\bibitem{Chaudhuri:2013xoa}
A.~Chaudhuri, W.~Grimus and B.~Mukhopadhyaya, \emph{{Doubly charged scalar
  decays in a type II seesaw scenario with two Higgs triplets}},
  \href{https://doi.org/10.1007/JHEP02(2014)060}{\emph{JHEP} {\bfseries 02}
  (2014) 060}, [\href{https://arxiv.org/abs/1305.5761}{{\ttfamily 1305.5761}}].

\bibitem{Chaudhuri:2016rwo}
A.~Chaudhuri and B.~Mukhopadhyaya, \emph{{CP -violating phase in a two Higgs
  triplet scenario: Some phenomenological implications}},
  \href{https://doi.org/10.1103/PhysRevD.93.093003}{\emph{Phys. Rev.}
  {\bfseries D93} (2016) 093003},
  [\href{https://arxiv.org/abs/1602.07846}{{\ttfamily 1602.07846}}].

\bibitem{Golling:2016gvc}
T.~Golling et~al., \emph{{Physics at a 100 TeV pp collider: beyond the Standard
  Model phenomena}},
  \href{https://doi.org/10.23731/CYRM-2017-003.441}{\emph{CERN Yellow Report}
  (2017) 441--634}, [\href{https://arxiv.org/abs/1606.00947}{{\ttfamily
  1606.00947}}].

\bibitem{Tang:2015qga}
J.~Tang et~al., \emph{{Concept for a Future Super Proton-Proton Collider}},
  \href{https://arxiv.org/abs/1507.03224}{{\ttfamily 1507.03224}}.

\bibitem{Arkani-Hamed:2015vfh}
N.~Arkani-Hamed, T.~Han, M.~Mangano and L.-T. Wang, \emph{{Physics
  opportunities of a 100 TeV proton–proton collider}},
  \href{https://doi.org/10.1016/j.physrep.2016.07.004}{\emph{Phys. Rept.}
  {\bfseries 652} (2016) 1--49},
  [\href{https://arxiv.org/abs/1511.06495}{{\ttfamily 1511.06495}}].

\bibitem{Olive:2016xmw}
{\scshape Particle Data Group} collaboration, C.~Patrignani et~al.,
  \emph{{Review of Particle Physics}},
  \href{https://doi.org/10.1088/1674-1137/40/10/100001}{\emph{Chin. Phys.}
  {\bfseries C40} (2016) 100001}.

\bibitem{Chun:2012jw}
E.~J. Chun, H.~M. Lee and P.~Sharma, \emph{{Vacuum Stability, Perturbativity,
  EWPD and Higgs-to-diphoton rate in Type II Seesaw Models}},
  \href{https://doi.org/10.1007/JHEP11(2012)106}{\emph{JHEP} {\bfseries 11}
  (2012) 106}, [\href{https://arxiv.org/abs/1209.1303}{{\ttfamily 1209.1303}}].

\bibitem{Baak:2014ora}
{\scshape Gfitter Group} collaboration, M.~Baak, J.~Cúth, J.~Haller,
  A.~Hoecker, R.~Kogler, K.~Mönig et~al., \emph{{The global electroweak fit at
  NNLO and prospects for the LHC and ILC}},
  \href{https://doi.org/10.1140/epjc/s10052-014-3046-5}{\emph{Eur. Phys. J.}
  {\bfseries C74} (2014) 3046},
  [\href{https://arxiv.org/abs/1407.3792}{{\ttfamily 1407.3792}}].

\bibitem{Frampton:2002yf}
P.~H. Frampton, S.~L. Glashow and D.~Marfatia, \emph{{Zeroes of the neutrino
  mass matrix}},
  \href{https://doi.org/10.1016/S0370-2693(02)01817-8}{\emph{Phys. Lett.}
  {\bfseries B536} (2002) 79--82},
  [\href{https://arxiv.org/abs/hep-ph/0201008}{{\ttfamily hep-ph/0201008}}].

\bibitem{Xing:2002ta}
Z.-z. Xing, \emph{{Texture zeros and Majorana phases of the neutrino mass
  matrix}}, \href{https://doi.org/10.1016/S0370-2693(02)01354-0}{\emph{Phys.
  Lett.} {\bfseries B530} (2002) 159--166},
  [\href{https://arxiv.org/abs/hep-ph/0201151}{{\ttfamily hep-ph/0201151}}].

\bibitem{Xing:2002ap}
Z.-z. Xing, \emph{{A Full determination of the neutrino mass spectrum from two
  zero textures of the neutrino mass matrix}},
  \href{https://doi.org/10.1016/S0370-2693(02)02062-2}{\emph{Phys. Lett.}
  {\bfseries B539} (2002) 85--90},
  [\href{https://arxiv.org/abs/hep-ph/0205032}{{\ttfamily hep-ph/0205032}}].

\bibitem{Honda:2003pg}
M.~Honda, S.~Kaneko and M.~Tanimoto, \emph{{Prediction and its stability in
  neutrino mass matrix with two zeros}},
  \href{https://doi.org/10.1088/1126-6708/2003/09/028}{\emph{JHEP} {\bfseries
  09} (2003) 028}, [\href{https://arxiv.org/abs/hep-ph/0303227}{{\ttfamily
  hep-ph/0303227}}].

\bibitem{Guo:2003cc}
W.-l. Guo and Z.-z. Xing, \emph{{Calculable CP violating phases in the minimal
  seesaw model of leptogenesis and neutrino mixing}},
  \href{https://doi.org/10.1016/j.physletb.2003.12.043}{\emph{Phys. Lett.}
  {\bfseries B583} (2004) 163--172},
  [\href{https://arxiv.org/abs/hep-ph/0310326}{{\ttfamily hep-ph/0310326}}].

\bibitem{Honda:2004qh}
M.~Honda, S.~Kaneko and M.~Tanimoto, \emph{{Seesaw enhancement of bilarge
  mixing in two zero textures}},
  \href{https://doi.org/10.1016/j.physletb.2004.03.093}{\emph{Phys. Lett.}
  {\bfseries B593} (2004) 165--174},
  [\href{https://arxiv.org/abs/hep-ph/0401059}{{\ttfamily hep-ph/0401059}}].

\bibitem{Goswami:2008rt}
S.~Goswami and A.~Watanabe, \emph{{Minimal Seesaw Textures with Two Heavy
  Neutrinos}}, \href{https://doi.org/10.1103/PhysRevD.79.033004}{\emph{Phys.
  Rev.} {\bfseries D79} (2009) 033004},
  [\href{https://arxiv.org/abs/0807.3438}{{\ttfamily 0807.3438}}].

\bibitem{Choubey:2008tb}
S.~Choubey, W.~Rodejohann and P.~Roy, \emph{{Phenomenological consequences of
  four zero neutrino Yukawa textures}},
  \href{https://doi.org/10.1016/j.nuclphysb.2009.04.021,
  10.1016/j.nuclphysb.2008.09.031}{\emph{Nucl. Phys.} {\bfseries B808} (2009)
  272--291}, [\href{https://arxiv.org/abs/0807.4289}{{\ttfamily 0807.4289}}].

\bibitem{Goswami:2009bd}
S.~Goswami, S.~Khan and W.~Rodejohann, \emph{{Minimal Textures in Seesaw Mass
  Matrices and their low and high Energy Phenomenology}},
  \href{https://doi.org/10.1016/j.physletb.2009.08.056}{\emph{Phys. Lett.}
  {\bfseries B680} (2009) 255--262},
  [\href{https://arxiv.org/abs/0905.2739}{{\ttfamily 0905.2739}}].

\bibitem{Fritzsch:2011qv}
H.~Fritzsch, Z.-z. Xing and S.~Zhou, \emph{{Two-zero Textures of the Majorana
  Neutrino Mass Matrix and Current Experimental Tests}},
  \href{https://doi.org/10.1007/JHEP09(2011)083}{\emph{JHEP} {\bfseries 09}
  (2011) 083}, [\href{https://arxiv.org/abs/1108.4534}{{\ttfamily 1108.4534}}].

\bibitem{Ghosh:2012pw}
M.~Ghosh, S.~Goswami and S.~Gupta, \emph{{Two Zero Mass Matrices and Sterile
  Neutrinos}}, \href{https://doi.org/10.1007/JHEP04(2013)103}{\emph{JHEP}
  {\bfseries 04} (2013) 103},
  [\href{https://arxiv.org/abs/1211.0118}{{\ttfamily 1211.0118}}].

\bibitem{Liao:2015hya}
J.~Liao, D.~Marfatia and K.~Whisnant, \emph{{Neutrino seesaw mechanism with
  texture zeros}},
  \href{https://doi.org/10.1016/j.nuclphysb.2015.09.020}{\emph{Nucl. Phys.}
  {\bfseries B900} (2015) 449--476},
  [\href{https://arxiv.org/abs/1508.07364}{{\ttfamily 1508.07364}}].

\bibitem{Kitabayashi:2015jdj}
T.~Kitabayashi and M.~Yasuè, \emph{{Formulas for flavor neutrino masses and
  their application to texture two zeros}},
  \href{https://doi.org/10.1103/PhysRevD.93.053012}{\emph{Phys. Rev.}
  {\bfseries D93} (2016) 053012},
  [\href{https://arxiv.org/abs/1512.00913}{{\ttfamily 1512.00913}}].

\bibitem{Lamprea:2016egz}
J.~M. Lamprea and E.~Peinado, \emph{{Seesaw scale discrete dark matter and
  two-zero texture Majorana neutrino mass matrices}},
  \href{https://doi.org/10.1103/PhysRevD.94.055007}{\emph{Phys. Rev.}
  {\bfseries D94} (2016) 055007},
  [\href{https://arxiv.org/abs/1603.02190}{{\ttfamily 1603.02190}}].

\bibitem{Grimus:2004az}
W.~Grimus and L.~Lavoura, \emph{{On a model with two zeros in the neutrino mass
  matrix}}, \href{https://doi.org/10.1088/0954-3899/31/7/014}{\emph{J. Phys.}
  {\bfseries G31} (2005) 693--702},
  [\href{https://arxiv.org/abs/hep-ph/0412283}{{\ttfamily hep-ph/0412283}}].

\bibitem{Beringer:1900zz}
{\scshape Particle Data Group} collaboration, J.~Beringer et~al., \emph{{Review
  of Particle Physics (RPP)}},
  \href{https://doi.org/10.1103/PhysRevD.86.010001}{\emph{Phys. Rev.}
  {\bfseries D86} (2012) 010001}.

\bibitem{Fogli:2012ua}
G.~L. Fogli, E.~Lisi, A.~Marrone, D.~Montanino, A.~Palazzo and A.~M. Rotunno,
  \emph{{Global analysis of neutrino masses, mixings and phases: entering the
  era of leptonic CP violation searches}},
  \href{https://doi.org/10.1103/PhysRevD.86.013012}{\emph{Phys. Rev.}
  {\bfseries D86} (2012) 013012},
  [\href{https://arxiv.org/abs/1205.5254}{{\ttfamily 1205.5254}}].

\bibitem{An:2012eh}
{\scshape Daya Bay} collaboration, F.~P. An et~al., \emph{{Observation of
  electron-antineutrino disappearance at Daya Bay}},
  \href{https://doi.org/10.1103/PhysRevLett.108.171803}{\emph{Phys. Rev. Lett.}
  {\bfseries 108} (2012) 171803},
  [\href{https://arxiv.org/abs/1203.1669}{{\ttfamily 1203.1669}}].

\bibitem{Ahn:2012nd}
{\scshape RENO} collaboration, J.~K. Ahn et~al., \emph{{Observation of Reactor
  Electron Antineutrino Disappearance in the RENO Experiment}},
  \href{https://doi.org/10.1103/PhysRevLett.108.191802}{\emph{Phys. Rev. Lett.}
  {\bfseries 108} (2012) 191802},
  [\href{https://arxiv.org/abs/1204.0626}{{\ttfamily 1204.0626}}].

\bibitem{Arbabifar:2012bd}
F.~Arbabifar, S.~Bahrami and M.~Frank, \emph{{Neutral Higgs Bosons in the Higgs
  Triplet Model with nontrivial mixing}},
  \href{https://doi.org/10.1103/PhysRevD.87.015020}{\emph{Phys. Rev.}
  {\bfseries D87} (2013) 015020},
  [\href{https://arxiv.org/abs/1211.6797}{{\ttfamily 1211.6797}}].

\bibitem{Kanemura:2012rs}
S.~Kanemura and K.~Yagyu, \emph{{Radiative corrections to electroweak
  parameters in the Higgs triplet model and implication with the recent Higgs
  boson searches}},
  \href{https://doi.org/10.1103/PhysRevD.85.115009}{\emph{Phys. Rev.}
  {\bfseries D85} (2012) 115009},
  [\href{https://arxiv.org/abs/1201.6287}{{\ttfamily 1201.6287}}].

\bibitem{Akeroyd:2012ms}
A.~G. Akeroyd and S.~Moretti, \emph{{Enhancement of H to gamma gamma from
  doubly charged scalars in the Higgs Triplet Model}},
  \href{https://doi.org/10.1103/PhysRevD.86.035015}{\emph{Phys. Rev.}
  {\bfseries D86} (2012) 035015},
  [\href{https://arxiv.org/abs/1206.0535}{{\ttfamily 1206.0535}}].

\bibitem{Dev:2013ff}
P.~S. Bhupal~Dev, D.~K. Ghosh, N.~Okada and I.~Saha, \emph{{125 GeV Higgs Boson
  and the Type-II Seesaw Model}},
  \href{https://doi.org/10.1007/JHEP03(2013)150,
  10.1007/JHEP05(2013)049}{\emph{JHEP} {\bfseries 03} (2013) 150},
  [\href{https://arxiv.org/abs/1301.3453}{{\ttfamily 1301.3453}}].

\bibitem{Das:2016bir}
D.~Das and A.~Santamaria, \emph{{Updated scalar sector constraints in the Higgs
  triplet model}},
  \href{https://doi.org/10.1103/PhysRevD.94.015015}{\emph{Phys. Rev.}
  {\bfseries D94} (2016) 015015},
  [\href{https://arxiv.org/abs/1604.08099}{{\ttfamily 1604.08099}}].

\bibitem{Sirunyan:2018ouh}
{\scshape CMS} collaboration, A.~M. Sirunyan et~al., \emph{{Measurements of
  Higgs boson properties in the diphoton decay channel in proton-proton
  collisions at $\sqrt{s} =$ 13 TeV}},
  \href{https://arxiv.org/abs/1804.02716}{{\ttfamily 1804.02716}}.

\bibitem{Alwall:2014hca}
J.~Alwall, R.~Frederix, S.~Frixione, V.~Hirschi, F.~Maltoni, O.~Mattelaer
  et~al., \emph{{The automated computation of tree-level and next-to-leading
  order differential cross sections, and their matching to parton shower
  simulations}}, \href{https://doi.org/10.1007/JHEP07(2014)079}{\emph{JHEP}
  {\bfseries 07} (2014) 079},
  [\href{https://arxiv.org/abs/1405.0301}{{\ttfamily 1405.0301}}].

\bibitem{Alloul:2013bka}
A.~Alloul, N.~D. Christensen, C.~Degrande, C.~Duhr and B.~Fuks,
  \emph{{FeynRules 2.0 - A complete toolbox for tree-level phenomenology}},
  \href{https://doi.org/10.1016/j.cpc.2014.04.012}{\emph{Comput. Phys. Commun.}
  {\bfseries 185} (2014) 2250--2300},
  [\href{https://arxiv.org/abs/1310.1921}{{\ttfamily 1310.1921}}].

\bibitem{Pumplin:2002vw}
J.~Pumplin, D.~R. Stump, J.~Huston, H.~L. Lai, P.~M. Nadolsky and W.~K. Tung,
  \emph{{New generation of parton distributions with uncertainties from global
  QCD analysis}},
  \href{https://doi.org/10.1088/1126-6708/2002/07/012}{\emph{JHEP} {\bfseries
  07} (2002) 012}, [\href{https://arxiv.org/abs/hep-ph/0201195}{{\ttfamily
  hep-ph/0201195}}].

\bibitem{ATLAS:2016pbt}
{\scshape ATLAS} collaboration, T.~A. collaboration, \emph{{Search for
  doubly-charged Higgs bosons in same-charge electron pair final states using
  proton-proton collisions at $\sqrt{s}=13\,\mathrm{TeV}$ with the ATLAS
  detector}}, ATLAS-CONF-2016-051.

\bibitem{Campbell:2012dh}
J.~M. Campbell and R.~K. Ellis, \emph{{$t \bar{t} W^{+-}$ production and decay
  at NLO}}, \href{https://doi.org/10.1007/JHEP07(2012)052}{\emph{JHEP}
  {\bfseries 07} (2012) 052},
  [\href{https://arxiv.org/abs/1204.5678}{{\ttfamily 1204.5678}}].

\bibitem{Lazopoulos:2008de}
A.~Lazopoulos, T.~McElmurry, K.~Melnikov and F.~Petriello,
  \emph{{Next-to-leading order QCD corrections to $t \bar{t} Z$ production at
  the LHC}}, \href{https://doi.org/10.1016/j.physletb.2008.06.073}{\emph{Phys.
  Lett.} {\bfseries B666} (2008) 62--65},
  [\href{https://arxiv.org/abs/0804.2220}{{\ttfamily 0804.2220}}].

\bibitem{Frixione:2015zaa}
S.~Frixione, V.~Hirschi, D.~Pagani, H.~S. Shao and M.~Zaro, \emph{{Electroweak
  and QCD corrections to top-pair hadroproduction in association with heavy
  bosons}}, \href{https://doi.org/10.1007/JHEP06(2015)184}{\emph{JHEP}
  {\bfseries 06} (2015) 184},
  [\href{https://arxiv.org/abs/1504.03446}{{\ttfamily 1504.03446}}].

\bibitem{Sjostrand:2006za}
T.~Sjostrand, S.~Mrenna and P.~Z. Skands, \emph{{PYTHIA 6.4 Physics and
  Manual}}, \href{https://doi.org/10.1088/1126-6708/2006/05/026}{\emph{JHEP}
  {\bfseries 05} (2006) 026},
  [\href{https://arxiv.org/abs/hep-ph/0603175}{{\ttfamily hep-ph/0603175}}].

\bibitem{deFavereau:2013fsa}
{\scshape DELPHES 3} collaboration, J.~de~Favereau, C.~Delaere, P.~Demin,
  A.~Giammanco, V.~Lemaître, A.~Mertens et~al., \emph{{DELPHES 3, A modular
  framework for fast simulation of a generic collider experiment}},
  \href{https://doi.org/10.1007/JHEP02(2014)057}{\emph{JHEP} {\bfseries 02}
  (2014) 057}, [\href{https://arxiv.org/abs/1307.6346}{{\ttfamily 1307.6346}}].

\bibitem{Cacciari:2008gp}
M.~Cacciari, G.~P. Salam and G.~Soyez, \emph{{The Anti-k(t) jet clustering
  algorithm}}, \href{https://doi.org/10.1088/1126-6708/2008/04/063}{\emph{JHEP}
  {\bfseries 04} (2008) 063},
  [\href{https://arxiv.org/abs/0802.1189}{{\ttfamily 0802.1189}}].

\bibitem{Conte:2012fm}
E.~Conte, B.~Fuks and G.~Serret, \emph{{MadAnalysis 5, A User-Friendly
  Framework for Collider Phenomenology}},
  \href{https://doi.org/10.1016/j.cpc.2012.09.009}{\emph{Comput. Phys. Commun.}
  {\bfseries 184} (2013) 222--256},
  [\href{https://arxiv.org/abs/1206.1599}{{\ttfamily 1206.1599}}].

\bibitem{ATLAS:2014pla}
{\scshape ATLAS} collaboration, T.~A. collaboration, \emph{{Calibration of the
  performance of $b$-tagging for $c$ and light-flavour jets in the 2012 ATLAS
  data}}, ATLAS-CONF-2014-046.

\bibitem{ATLAS:2016rin}
{\scshape ATLAS} collaboration, T.~A. collaboration, \emph{{Measurement of the
  $W^+W^-$ production cross section in $pp$ collisions at a centre-of-mass
  energy of $\sqrt{s}=13$ TeV with the ATLAS experiment}}, ATLAS-CONF-2016-090.

\bibitem{ATLAS:2016iqc}
{\scshape ATLAS} collaboration, T.~A. collaboration, \emph{{Electron efficiency
  measurements with the ATLAS detector using the 2015 LHC proton-proton
  collision data}}, ATLAS-CONF-2016-024.

\bibitem{Aad:2016jkr}
{\scshape ATLAS} collaboration, G.~Aad et~al., \emph{{Muon reconstruction
  performance of the ATLAS detector in proton–proton collision data at
  $\sqrt{s}$ =13 TeV}},
  \href{https://doi.org/10.1140/epjc/s10052-016-4120-y}{\emph{Eur. Phys. J.}
  {\bfseries C76} (2016) 292},
  [\href{https://arxiv.org/abs/1603.05598}{{\ttfamily 1603.05598}}].


\bibitem{Mangano:2016jyj}
M.~L. Mangano et~al., \emph{{Physics at a 100 TeV pp Collider: Standard Model
  Processes}}, \href{https://doi.org/10.23731/CYRM-2017-003.1}{\emph{CERN
  Yellow Report} (2017) 1--254},
  [\href{https://arxiv.org/abs/1607.01831}{{\ttfamily 1607.01831}}].

\end{thebibliography}

\providecommand{\href}[2]{#2}\begingroup\raggedright\endgroup

\end{document}